\documentclass[nofootinbib,aps,prd,reprint,superscriptaddress,showkeys]{revtex4-2}

\usepackage[utf8]{inputenc} 
\usepackage{graphicx,overpic,mathtools}
\usepackage{amsthm,amsmath,amssymb,hyperref}
\usepackage{braket,bm,bbm,setspace}
\usepackage[normalem]{ulem} 
\usepackage{physics}
\usepackage{float}
\usepackage[makeroom]{cancel}
\usepackage[english]{babel}
\usepackage{xcolor}
\usepackage{tensor}
\graphicspath{ {images/} }
\addto\captionsspanish{}
\hypersetup{
	colorlinks=true,
	pdfborder={0 0 0},
	citecolor=purple,
	linkcolor=blue,
	filecolor=blue,
	urlcolor=blue,
}
\usepackage{subfigure}

\usepackage{orcidlink}

\usepackage{multirow}

\usepackage{tikz}
\usetikzlibrary{quantikz2}

\makeatletter
\def\th@plain{
	\thm@notefont{}
	\itshape
}
\def\th@definition{
	\thm@notefont{}
	\normalfont
}
\makeatother

\begin{document}

\newcommand{\ci}{\text{i}}
\newcommand{\identity}{\mathbbm{1}}
\newcommand{\cs}{0.2cm}
\newcommand{\css}{0.3cm}
\newcommand{\rs}{0.3cm}
\newcommand{\reference}{[CITE] }
\newcommand{\CNOT}{\text{CNOT}}
\newcommand{\eqp}{\sim}
\newcommand{\nT}{M}
\newcommand{\nR}{\mathcal{R}}
\newcommand{\nt}{m}
\newcommand{\QASM}{Aer }

\newcommand{\greater}{greater }
\newcommand{\larger}{larger }

\newcommand{\apriori}{\textit{a priori} }
\newcommand{\Apriori}{\textit{A priori} }

\title{Reverse Delegated Training and Private Inference via\\Perfectly-Secure Quantum Homomorphic Encryption}
%\author{Sergio A. Ortega}
\author{Sergio A. Ortega \orcidlink{0000-0002-8237-7711}}
\email{sergioan@ucm.es}
\affiliation{Departamento de Física Teórica, Universidad Complutense de Madrid, 28040 Madrid, Spain}
%\author{Miguel A. Martin-Delgado}
\author{Miguel A. Martin-Delgado \orcidlink{0000-0003-2746-5062}}
\email{mardel@ucm.es}
\affiliation{Departamento de Física Teórica, Universidad Complutense de Madrid, 28040 Madrid, Spain}
\affiliation{CCS-Center for Computational Simulation, Universidad Politécnica de Madrid, 28660 Boadilla del Monte, Madrid, Spain}

\begin{abstract}
	{Quantum machine learning in cloud environments requires protecting sensitive data while enabling remote computation. Here we demonstrate the first realistic implementations of a perfectly-secure quantum homomorphic encryption (QHE) scheme applied to quantum neural networks (QNN). Using efficient Clifford+$T$ decomposition, we implement quantum convolutional neural networks for two complementary scenarios: (i) reverse delegated training, where encrypted data from multiple providers trains a user's network via federated aggregation; (ii) private inference, where users process encrypted data with remote quantum networks. Moreover, analysis of server circuit privacy reveals probabilistic model protection through Pauli gate concealment. These results establish perfectly-secure QHE as a practical framework for multi-party quantum machine learning.}
\end{abstract}

\keywords{Quantum Machine Learning, Quantum Neural Networks, Quantum Homomorphic Encryption.}

\maketitle

\onecolumngrid

\section{Introduction}\label{Introduction}

Artificial intelligence is a revolutionary paradigm transforming science and society \cite{Deep_learning}, with emerging applications in quantum computing promising computational advantages over classical methods \cite{QASUML,Paparo3,Power_QNN}. The integration of quantum computing with machine learning creates the field of quantum machine learning \cite{QML,QML-2,QASUML,Power_QNN}, where quantum neural networks (QNN) \cite{QNN} are expected to provide meaningful advantages even in the Noisy Intermediate-Scale Quantum (NISQ) era \cite{Preskil,NISQA,VQA,PQCML}.\\

Cloud quantum computing represents a growing paradigm where users lacking sufficient quantum resources delegate computation to remote quantum servers. This enables widespread access to quantum algorithms, particularly quantum neural networks. However, such scenarios introduce critical privacy concerns: sensitive data (medical, financial, proprietary) used in training and inference must remain confidential from both the server infrastructure and any potential adversaries.\\

Quantum homomorphic encryption (QHE) provides an ideal framework for data protection, enabling quantum algorithms to operate on encrypted data without decryption at the server \cite{QFHE_def,T_interactions,Broadbent,No_go_result,Enhanced_no_go}. A promising QHE scheme \cite{Liang} offers significant advantages over prior approaches: (i) perfect information-theoretic security (rather than computational security), ensuring zero data leakage even against computationally unbounded adversaries; (ii) non-interactive evaluation, eliminating continuous communication requirements; (iii) $\mathcal{F}$-homomorphic capability for universal quantum computation; (iv) quasi-compactness with complexity scaling linearly with non-Clifford T-gates, enabling efficient implementation for polynomial-T-gate-scaling algorithms.\\

Whereas works in the literature have explored the use of QHE for QNN, the schemes used usually provide a computational security \cite{QNN-QHE-1,QNN-QHE-2,QNN-QHE-3,QNN-QHE-5,QNN-QHE-6,QNN-QHE-7}. Therefore, in this work we explore the implementation of the perfectly-security scheme \cite{Liang} in the field of quantum machine learning to obtain perfect security, so that there is no data leakage even if a malicious user has infinite computational power.\\

Moreover, works are mainly focused on the use of quantum homomorphic encryption for clients with low quantum capabilities, so that they ask a server to run their networks while protecting their data, both for training and inference \cite{QNN-QHE-1,QNN-QHE-2,QNN-QHE-3,QNN-QHE-5,QNN-QHE-6,QNN-QHE-7}. Although the QHE scheme we explore can address these conventional scenarios, our focus here differs. However, in this work we focus on other point of view, where even if a client had enough power to run the network by itself, another party is necessarily involved, and then can take advantage of QHE. Instead of a single user owing both the data and the QNN, we focus on cases where a user has the data and wants to keep it secret from another user who owns the network. Thus, we propose the following two applications of this scheme:

\begin{itemize}
	\item Reverse Delegated Training.
	
	In the usual delegated training a user owns both the data and the quantum network model. However, lacking enough quantum computational power, asks a server to execute the network during the training process. To do so, this user sends the QNN circuit model with the encrypted data to the server, so that the server can run the network on them, and return the results to the user. Instead of this canonical scenario, we propose the case where a user has a QNN to train, and also has the quantum resources to execute it. However, now the user lacks of data for the training. Therefore, this user asks another party to supply the data, which sends it in an encrypted manner. Thus, in this case the owner of the data provides a service to the user that executes the network, contrary to the usual delegated scenario.
	
	\item Private Inference.
	
	In the usual delegated inference scenario a user has both the data and the QNN model. Again, it lacks enough quantum resources to execute, so that asks a server to run it, sending the encrypted data along with the network circuit. In our private inference scenario instead, we propose a user which only has data, and wants to make inference on them with a quantum neural network that a server keeps secret. Therefore, the user sends the encrypted data to the server, and this runs the private network on them.
\end{itemize}

In both scenarios, the necessity of involving another party stems not from limited quantum resources---as in typical cases-but from lacking either data or the network model.\\

In this work we explore the application of these two frameworks with instances of quantum convolutional neural networks (QCNN) \cite{CNN}. We also prove that they can be implemented efficiently, by explicitly calculating the complexity in terms of $T$ gates. Moreover, since the owner of the network wants to keep its weights secret, it is essential that the user providing the data do not obtain such information during the decryption process. We also study this possibility for this scheme \cite{Liang}, which is currently an open problem \cite{Homomorphic}.\\

Furthermore, we provide realistic vs. proof-of-concept simulations in this work. Previous QHE simulations \cite{QHE_grover,Pablo_Grover,Homomorphic} implemented the QHE scheme \cite{Liang} on algorithms where sensitive information was embedded in algorithm structure \cite{Grover,Szegedy}. Our work represents the first realistic QHE-QNN simulations where sensitive information resides solely in user-provided quantum states, matching practical scenarios where clients provide data and servers provide algorithms.\\

This paper is structured as follows. In Section \ref{sec:QHE} we review the formulation of the perfectly-secure QHE scheme \cite{Liang}. In Section \ref{sec:QNN} we review the theory of quantum neural networks, with focus on QCNN architectures. In Section \ref{sec:QHE-QNN} we study the efficient implementation of the QCNN with QHE. In section \ref{sec:Training} we demonstrate reverse delegated training with federated aggregation. In section \ref{sec:Inference} we explore private inference with server circuit privacy analysis. Finally, we summarize and conclude in Section \ref{sec:Conclusions}.

\section{Quantum Homomorphic Encryption}\label{sec:QHE}

\subsection{Scheme description}

To describe the scheme for QHE \cite{Liang}, let us first define properly the two parties involved. We define as CLIENT to the party that creates the initial state $\left|\alpha\right>$ of the system and reads the final result $\left|\beta\right>$, whereas SERVER is the party that runs the quantum algorithm $U$ on the initial state. In our analysis, we use the terms `CLIENT' and `SERVER' in a purely functional sense within the QHE framework. We do this regardless of business models or payment arrangements, since quantum neural network applications---as discussed below---allow either party to functionally serve as the client depending on the use case. The QHE scheme consists of three main procedures, two of them performed by CLIENT and one by SERVER, and the quantum circuit performing the algorithm $U$ must be decomposed in Clifford+$T$ gates from the set $\mathcal{G} = \lbrace{X,Z,H,S,T,\CNOT\rbrace}$. Note that although the gates $H$ and $T$ are the only ones necessary for universal quantum computation, the gates $S = T^2$, $Z=S^2$, and $X=HZH$ are also introduced because at a logical level their homomorphic implementation is by far simpler than considering $T$ gates.\\

{\bfseries Step 1:} Client initializes a $n$-qubit system in the desired initial state $\left|\alpha\right>$ of the quantum algorithm. Then, CLIENT generates randomly a secret key composed of two bit strings of length $n$, and stores them in classical-bitstring variables $x$ and $z$:
\begin{equation}
	x = x_1x_2...x_n, \ \ \ z = z_1z_2...z_n.
\end{equation} 
CLIENT uses the keys to encrypt the system with quantum one-time pad (QOTP) encryption \cite{One_time_pad}. For doing so, CLIENT applies to each qubit $q_i$ the Pauli gates $X$ and $Z$ depending on the classical bits $x_i$ and $z_i$:
\begin{equation}
	\left|\alpha^{enc}\right> = \left[\bigotimes_{i=1}^n X_i^{x_i}Z_i^{z_i}\right] \left|\alpha\right>.
\end{equation}
Since each time the algorithm is repeated the initial encryption key is chosen at random, the state that SERVER receives is totally randomized. This so because if we select the encryption bitstrings at random and use them only once, applying this protocol to any arbitrary quantum state $\rho$ produces the totally mixed state \cite{One_time_pad}:
\begin{equation}
	\frac{1}{2^{2n}} \sum_{a,b \in \lbrace{0,1\rbrace}^{n}} \left[\bigotimes_{i=1}^n X_i^{a_i}Z_i^{b_i}\right] \rho \left[\bigotimes_{i=1}^n X_i^{a_i}Z_i^{b_i}\right]^\dagger = \frac{\mathbbm{1}_{2^n}}{2^n}.
\end{equation}
Therefore, QOTP provides perfect security to the QHE scheme, and SERVER cannot obtain any information about the state of CLIENT.\\

{\bfseries Step 2:} After receiving the encrypted state $\left|\alpha^{enc}\right>$ by CLIENT, SERVER runs homomorphically the circuit of the algorithm unitary $U$. To do so, For each gate $g$ in the original circuit, SERVER applies its corresponding homomorphic evaluation scheme. Whereas this evaluation consists trivially of applying the gate $g$ for Clifford gates, it is quite more complicated for the $T$ gates, as we show in the next section.\\

After each gate evaluation there is a change in the encrypting key of the system, so that the values stored in the classical bits $x_i$ and $z_i$ will have to be updated accordingly. Therefore, for each gate $g$ applied to qubit $q_i$, there is associated a key-updating function, denoted as $f_{g,i}$. Thus, SERVER must generate a sequence with the key updating functions accordingly to the sequence of gates in the quantum circuit, so that CLIENT can later decrypt the final state $\left|\beta^{enc}\right>$. These functions are shown in Table \ref{tab:gates}.\\

\begin{table}[htbp]
	\centering
	\caption{Homomorphic evaluation rules and corresponding key-updating functions for each gate in the Clifford+T set. Note that for the Pauli gates $X$ and $Z$ the key is not updated, so that there is no a key-updating function. For the $T$ gate, $\widetilde{T}$ represents the evaluation scheme in Figure \ref{F:T-evaluation}, and the key-updating function requires quantum operations to obtain the values $r_a$ and $r_b$. The symbol $\eqp$ denotes equality up to a global phase.}
	\begin{tabular}{l|l}
		\hline
		Homomorphic evaluation & Key-updating function\\
		\hline
		$X_i X_i^aZ_i^b\left|\phi\right> \eqp X_i^aZ_i^b X_i \left|\phi\right>$ & $---$\\
		$Z_i X_i^aZ_i^b\left|\phi\right> \eqp X_i^aZ_i^b Z_i \left|\phi\right>$ & $---$ \\
		$H_i X_i^aZ_i^b\left|\phi\right> = X_i^bZ_i^a H_i \left|\phi\right>$ & $f_{H,i}(x_i,z_i) = (z_i,x_i)$ \\
		$S_i X_i^aZ_i^b\left|\phi\right> \eqp X_i^aZ_i^{a \oplus b} S_i \left|\phi\right>$ & $f_{S,i}(x_i,z_i) = (x_i,x_i \oplus z_i)$ \\
		$\CNOT_{ij} X_i^{a_i}Z_i^{b_i}X_j^{a_j}Z_j^{b_j}\left|\phi\right> = X_i^{a_i}Z_i^{b_i \oplus b_j}X_j^{a_i \oplus a_j}Z_j^{b_j} \CNOT \left|\phi\right>$ & $f_{C,ij}((x_i,z_i),(x_j,z_j)) = ((x_i,z_i \oplus z_j),(x_i \oplus x_j,z_j))$\\
		\hline
		&\\
		$\widetilde{T}_{i;\nt} X_i^aZ_i^b\left|\phi\right> \eqp X^{a\oplus r_{a_\nt}}Z^{a\oplus b \oplus r_{b_\nt}}T\left|\phi\right>$ & $f_{T,i;\nt}(x_i,z_i) = (x_i \oplus r_{a_\nt},x_i \oplus z_i \oplus r_{b_\nt})$\\
		\hline
	\end{tabular}
	\label{tab:gates}%
\end{table}%

{\bfseries Step 3:} Since the final state received from SERVER is encrypted, CLIENT must decrypt it to obtain the correct result. However, CLIENT needs to update the encrypting keys first. To do so, for each function $f_{g,i}$ in the sequence of key-updating functions, the classical bits are updated:
\begin{equation}
	x_i,z_i \leftarrow f_{g,i}(x_i,z_i),
\end{equation}
\begin{equation}
	(x_i,z_i),(x_j,z_j) \leftarrow f_{C,ij}((x_i,z_i),(x_j,z_j)).
\end{equation}
At the end of the sequence, these classical bits contain the final encryption key, and CLIENT can decrypt the final state applying QOTP decryption:
\begin{equation}
	\left|\beta\right> = \left[\bigotimes_{i=1}^n X_i^{x_i}Z_i^{z_i}\right] \left|\beta^{enc}\right>.
\end{equation}

In most cases, as for example in the quantum neural networks we explore in this work, CLIENT wants to measure the final state $\left|\beta\right>$ in the computational basis. In this case, SERVER can measure directly $\left|\beta^{enc}\right>$ and send the resulting bitstring to CLIENT, which is encrypted with the key stored in classical bitstring $x$. Thus, CLIENT can decrypt directly the classical result with this key, using just classical XOR operations.

\subsection{Homomorphic Evaluation of $T$ Gates}

\begin{figure}[hbpt]
	\centering
	\makebox[0pt][c]{
	\subfigure[]{
	\begin{quantikz}[column sep = 0.18cm]
		\lstick[1]{$X^aZ^b\left|\phi\right>$} &\gate[1]{T}&&\swap[]{1} \slice{}&&&&& \rstick[1]{ $X^{a\oplus r_a}Z^{a\oplus b \oplus r_b}T\left|\phi\right>$}\\
		\lstick[2]{$\left|\Phi\right>_{00}$:}&  &  & \targX{}&\gate{S^a} & \ctrl{1} & \gate{H}& \meter{}&\setwiretype{c}\rstick[1]{$r_b$}\\
		&&&&&\targ{}&&\meter{}&\setwiretype{c}\rstick[1]{$r_a$}
	\end{quantikz}
	\label{F:T-evaluation}}
	\subfigure[]{\begin{quantikz}[column sep = 0.1cm]
		&\gate[1]{T}&&&&\swap[]{1}&\\
		\lstick[2]{Bell$_m$:}&\setwiretype{n}\left|0\right>&\setwiretype{q}&\gate{H} & \ctrl{1} & \targX{}&\\
		&\setwiretype{n}\left|0\right>&&\setwiretype{q}&\targ{}&&
	\end{quantikz}
	\label{F:T-Server}}
	\subfigure[]{\begin{quantikz}[column sep = 0.1cm]
		&&&&&&\\
		\lstick[2]{Bell$_m$:}&\gate{S^a} & \ctrl{1} & \gate{H}& \meter{}&\setwiretype{c}\rstick[1]{$r_b$}\\
		&&\targ{}&&\meter{}&\setwiretype{c}\rstick[1]{$r_a$}
	\end{quantikz}
	\label{F:T-Client}}
	}
	\caption{(a) Homomorphic evaluation scheme for a T gate using quantum teleportation. (b) Circuit corresponding to SERVER’s operations, including Bell-pair creation and swap. (c) CLIENT's decryption procedure, including a Bell measurement after the correction of the $S$ error.}
	\label{F:...}
\end{figure}
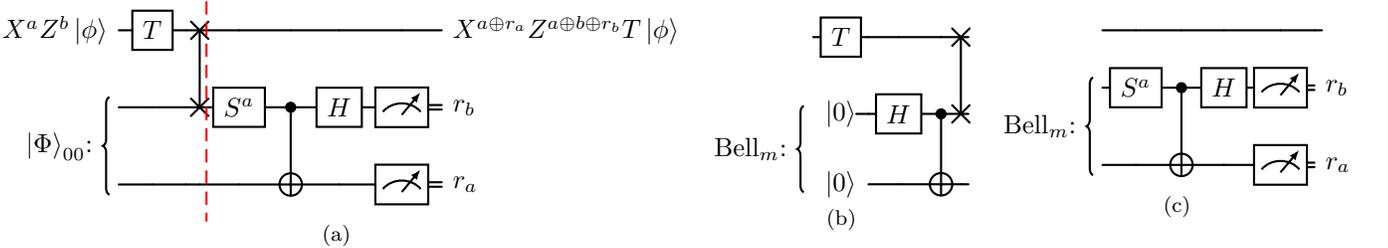

The homomorphic evaluation of Clifford gates is trivial, just applying the corresponding gate in the circuit. This is so because when Clifford gates are applied to any state encrypted with $X$ and $Z$ gates, they produce another state also encrypted with $X$ and $Z$ gates, as shown in Table \ref{tab:gates}. However, the case of a $T$ gate is more complex. Let $\left|\phi\right>$ be an arbitrary quantum state, as for example any intermediate state of the quantum algorithm. For the sake of simplicity let us suppose that the gate acts on the first qubit of $\left|\phi\right>$, which is encrypted with $x_1=a$ and $z_1=b$ in the first qubit, so that we have $X_1^{a}Z_1^{b}\left|\phi\right>$. Let us then apply the $T$ gate:
\begin{equation}
	T_1 X_1^aZ_1^b\left|\phi\right> \eqp (S_1^\dagger)^a X_1^aZ_1^{a \oplus b} T_1 \left|\phi\right>.
\end{equation}
The result $T\left|\phi\right>$ is now encrypted by $X$ and $Z$ gates with an updated key. However, there is also an additional phase error $(S^\dagger)^a$. The bit value $a$ is given by the encrypting-key variable $x_1$ before the application of the $T$ gate. However, this depends in turn on the initial encrypting key, which SERVER does not know, and thus is unable to correct this error by itself. If SERVER sent the qubit to CLIENT for correcting the error and turning it back, then the QHE scheme would be interactive. Instead, an evaluation scheme that makes use of quantum teleportation was introduced \cite{Liang}, which is shown in Figure \ref{F:T-evaluation}. Server initializes two ancilla qubits in the Bell state
\begin{equation}
	\left|\Phi\right>_{00} = \frac{1}{\sqrt{2}}\left|00\right> + \frac{1}{\sqrt{2}}\left|11\right>.
\end{equation}
After that, SERVER applies a swap gate between the problem qubit and the first qubit of the Bell pair. At this point, Server would send the two ancilla qubits to CLIENT, so that CLIENT can apply the gate $S^a$ to the problem qubit, and teleport it back to Server. Thus, the qubit of SERVER ends up with the phase error corrected. However, since the scheme is non-interactive, instead of communicating to SERVER the results of the measurements $r_a$ and $r_b$ to finish the teleportation protocol, these results are considered part of the updated encrypting key. Thus, the system of SERVER ends up in the state $X_1^{a\oplus r_a}Z_1^{a\oplus b \oplus r_b}T\left|\phi\right>$.

Furthermore, due to the principle of deferred measurements \cite{Nielsen}, CLIENT can postpone these measurements until SERVER has finished running the quantum circuit. Thus, SERVER can indeed wait until the end to send the ancilla qubits along with the main qubits of the system. Taking all this into account, the evaluation scheme for a $T$ gate corresponds to a creation of a Bell pair in two ancilla qubits using a Hadamard gate and a CNOT gate, and a swap, as shown in Figure \ref{F:T-Server}; and the key-updating procedure of CLIENT consists on applying the corresponding $S^a$ gate and the measurement in the Bell basis, as shown in Figure \ref{F:T-Client}.\\

Note that we need a different register of two qubits, from now on denoted as Bell registers, for each $T$ gate in the circuit. Let $\nT$ be the number of $T$ gates. Therefore, we have in the end $\nT$ Bell registers, which can be indexed by the subscript $\nt = 1,...,\nT$. For the decryption procedure related to each $T$ gate, CLIENT must measure the corresponding Bell register and update the classical keys, so that we can describe the whole process as a classical-quantum function indexed by both the qubit where the $T$ gate acts on and the corresponding index $\nt$:
\begin{equation}
	f_{T,i;\nt}(x_i,z_i) = (x_i \oplus r_{a_\nt},x_i \oplus z_i \oplus r_{b_\nt}),
\end{equation}
where the values $r_{a_\nt}$ and $r_{b_\nt}$ are the results of the quantum measurements of the $\nt$-th Bell register.

\subsection{Scheme Analysis}

Whereas for each Clifford gate the decryption procedure only involves classical operations, $T$ gates require of quantum operations and measurements. Moreover, each $T$ gate requires two ancilla qubits to be sent from SERVER to CLIENT. For this reason, the complexity of the scheme is usually measured in terms of $T$ gates, similarly as for quantum error correction schemes for fault-tolerant computation \cite{Error_correction,T-complexity,TFermion}. In the case of this QHE scheme, the number of quantum operations for decryption scales linearly with the number $\nT$ of $T$ gates, and for this reason it is $\nT$-quasicompact. Therefore, any quantum algorithm can be implemented efficiently provided that $\nT$ grows at most as a polynomial with the size of the problem \cite{Liang}.\\

Since the circuit is decomposed in a discrete set of universal gates, the depth may be so long that quantum error correction is required for a reliable implementation. Nevertheless, note that although CLIENT also needs quantum capabilities for the decryption procedure, all the gates involved are Clifford. Thus, whereas SERVER would need complex quantum error correction schemes for each $T$ gate applied in the circuit, CLIENT needs by far less quantum resources since Clifford gates can be corrected easier.\\

Regarding the creation of the initial state prior to the encryption, the circuit does not need to be decomposed on this universal set. Then, an implementation using parameterized rotation gates suitable for the NISQ era \cite{Preskil} may be enough, as is the case of quantum neural network circuits \cite{VQA}. Moreover, even in the case that the preparation of the initial state involves so many resources that CLIENT also needs to have good quantum capabilities, this QHE scheme can be justified in cases where the algorithm is owned by SERVER, and it does not want to share it with CLIENT. An example of this scenario is explored in Section \ref{sec:Inference}.

\section{Quantum Neural Networks}\label{sec:QNN}

From a mathematical point of view, a neural network is just a parameterized function $f(\mathbf{x},\mathbf{w})$, where $\mathbf{x}$ is a vector of data features, and $\mathbf{w}$ is a vector of parameters denoted as weights. In supervised machine learning the aim is to fit this function to a set $X$ of labeled data $(\mathbf{x},y)$, so that the fitted function can make predictions about new unseen data. Fitting this function corresponds to minimizing a loss function, which measures the difference between the output of the function and the true label $y$ of the data $\mathbf{x}$. This process of optimization is usually referred to as training, since it is as if the artificial intelligence learns by training with the data.\\

\begin{figure}[hbpt]
	\centering
	\makebox[0pt][c]{
	\includegraphics[width=0.7\linewidth]{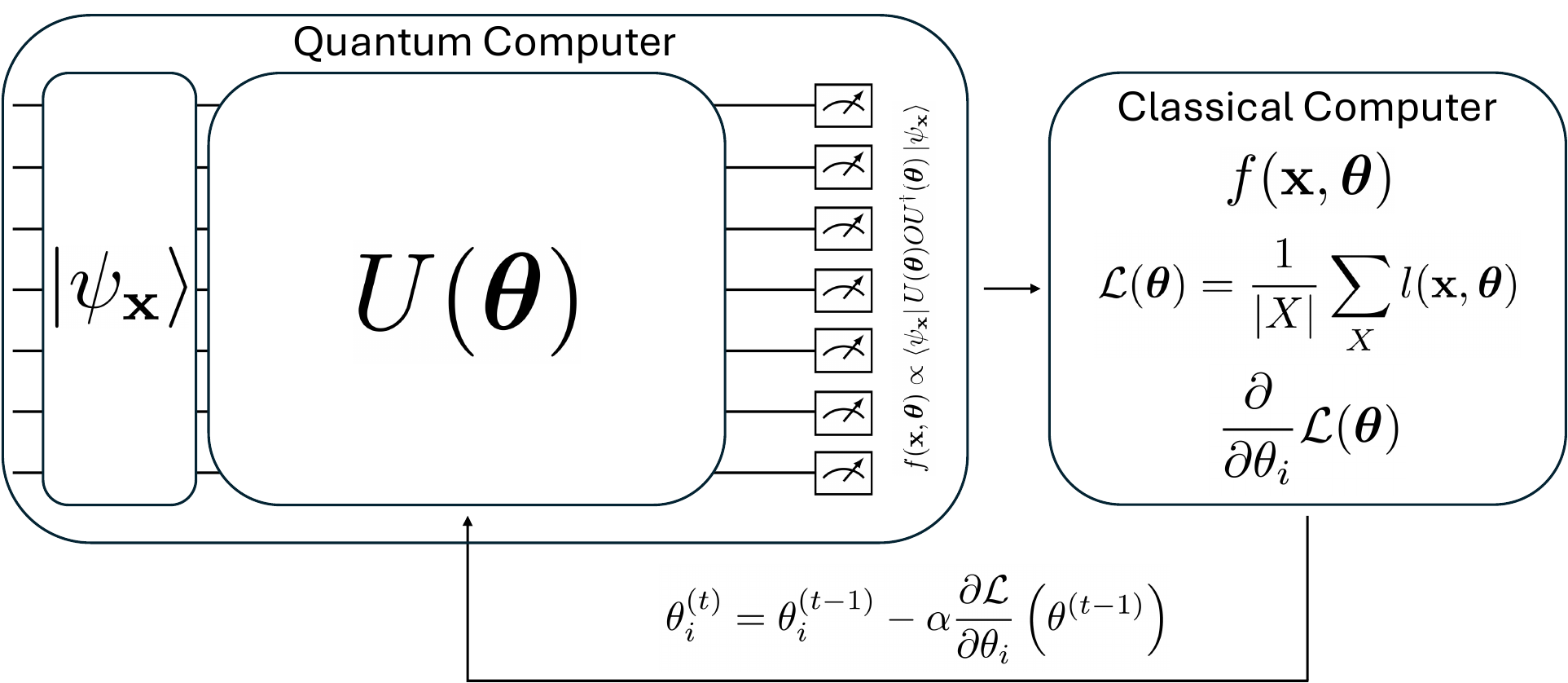}
	}
	\caption{General workflow of a quantum neural network training process. At each iteration, the initial state $\left|\psi_\mathbf{x}\right>$ representing the data is fed to the quantum network, processed by a parameterized quantum circuit $U(\boldsymbol{\theta})$ of the ansatz, and measured to produce classical outputs used for calculating the result $f(\mathbf{x},\boldsymbol{\theta})$. With this, the loss and its gradient are calculated, and finally the weights $\boldsymbol{\theta}$ are updated on a classical computer.}
	\label{F:QNN}
\end{figure}

In the context of quantum computing, the analogue quantum neural network corresponds to a quantum function $f(\mathbf{x},\boldsymbol{\theta})$, whose output is related to the expectation value of some observable $O$ for a parameterized quantum circuit (PQC). In this case the weights are usually denoted as $\boldsymbol{\theta}$ since they correspond to angles of rotation gates in the circuit. Regarding the data, each instance is represented as a quantum state $\left|\psi_\mathbf{x}\right>$, and the unitary operator of the ansatz PQC is represented as $U(\boldsymbol{\theta})$. Therefore:
\begin{equation}
	f(\mathbf{x},\boldsymbol{\theta}) \propto \left<\psi_\mathbf{x}\right|U(\boldsymbol{\theta})OU^\dagger(\boldsymbol{\theta})\left|\psi_\mathbf{x}\right>.
\end{equation}
Regarding the data, it can be pure quantum data, which comes for example as the output of some quantum algorithm, or is a state obtained from a quantum sensor \cite{QSVDD,Quantum_sensor}. However, it can also represent classical data, for which it is necessary to encode the classical data into a quantum state with a quantum feature map \cite{Schuld_feature_space}. For the sake of this work we are going to deal with the task of classifying classical data, and in the following sections we show how this data can be encoded, as well as the ansatz for the PQC, the loss function, and how it is optimized. A general overview of the process is represented in Figure \ref{F:QNN}.

\subsection{Data encoding}\label{sec:data}

Encoding the data into a quantum state requires a quantum circuit depending on the feature vector $\mathbf{x}$ that, when applied to the state $\left|0\right>$, produces the corresponding quantum data state $\left|\psi_\mathbf{x}\right>$ \cite{Schuld_feature_space,QML_book,QCNN}.\\

A naive form of encoding is qubit encoding, which corresponds to applying a Pauli rotation gate to each qubit, where the angles are given by the classical data vector. In the case of using $R_Y$ gates, the initial state is
\begin{equation}
	\left|\psi_\mathbf{x}\right> = \bigotimes_{i=1}^n\left[\cos\left(\frac{x_i}{2}\right)\left|0\right>+\sin\left(\frac{x_i}{2}\right)\left|1\right>\right].
\end{equation}
This form of encoding is very convenient because it just requires a rotation gate per qubit. However, the number of qubits grows linearly with the size of the vector data $\mathbf{x}$ since we require a qubit per feature.\\

Another form of encoding which exploits the exponential capability of quantum computers to store information is amplitude encoding. In this case each classical feature corresponds to an amplitude in the corresponding normalized quantum state:
\begin{equation}
	\left|\psi_\mathbf{x}\right> = \frac{1}{||\mathbf{x}||}\sum_{i=1}^{2^n}x_i\left|i\right>.
\end{equation}
There exist algorithms for creating the corresponding encoding circuit given the classical vector \cite{Initializer}, and with this encoding, only $n$ qubits are needed to represent a classical feature vector with $N=2^n$ components. However, in contrast to qubit encoding, where the circuit has a constant depth, the depth in this case grows as $\mathcal{O}(N)$. Therefore, using one or other encoding method depends of the quantum capabilities and the size of the data, and in the end, of the particular application.\\

While more complex feature maps exist (e.g., ZZ feature maps \cite{ZZ}), we focus on qubit and amplitude encoding. These methods are particularly suitable because their outputs directly represent classical data—accessible by computational-basis measurements alone. This direct representation makes them ideal for leveraging QHE-based data protection.

\subsection{Ansatz}

The core of the quantum neural network is a PQC denoted as ansatz, whose weights must be fitted to capture the relationship between the input data and its label. In this work we deal with quantum convolutional neural networks (QCNN). These have a structure inspired by the classical convolutional neural networks, which have shown good performance for tasks as image classification \cite{CNN}. Moreover, QCNN have a translation invariant structure which allows to capture not only local correlations, but also global correlations in the data, and have been proven to be free of barren plateaus \cite{No-barren}. Furthermore, they are suitable for the NISQ era, and therefore we expect that they can also be decomposed efficiently into the Clifford+$T$ set, which we explore in Section \ref{sec:T_count}.\\

\begin{figure}[hbpt]
	\centering
	\makebox[0pt][c]{
	\includegraphics[width=0.7\linewidth]{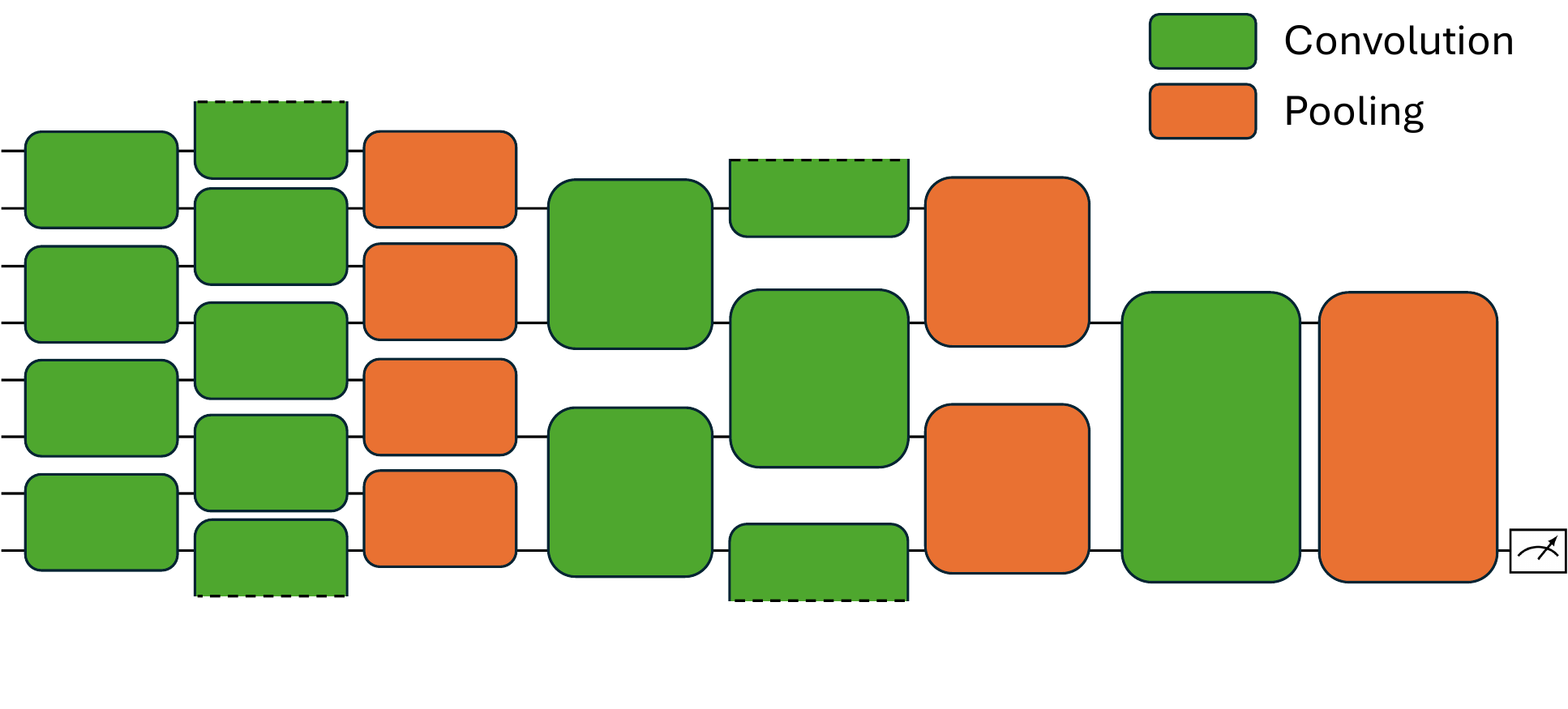}
	}
	\caption{Example of the structure of the quantum convolutional neural network ansatz for a system with eight qubits. In this case there are three layers of convolutional and pooling operations. After the pooling operations, the number of effective qubits is reduced by half. The dashed line in a convolutional unit means that it connects with the opposite qubit in the circuit.}
	\label{F:QCNN}
\end{figure}

Figure \ref{F:QCNN} illustrates the ansatz structure. Inspired by classical convolutional neural networks, the circuit comprises multiple layers of convolution and pooling operations. Each convolutional layer contains subunits---operators acting on pairs of qubits---that collectively capture multi-qubit correlations. The convolutional layer contains subunits that connect qubits in a cyclic chain, ensuring translation invariance. Pooling operators then act on non-overlapping consecutive qubit pairs, projecting out one qubit per pair. This halves the qubit count at each layer. After $\log_2(n)$ convolution-pooling iterations, a single qubit remains, which we measure to obtain the function output $f(\mathbf{x},\boldsymbol{\theta})$.\\

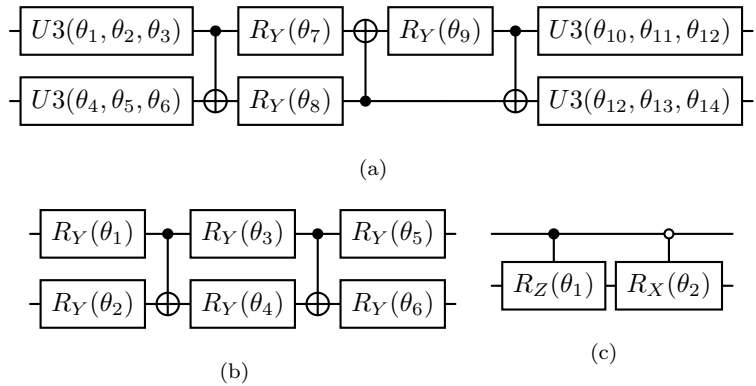
\begin{figure}[hbpt]
	\centering
	\subfigure[]{
		\begin{quantikz}[column sep = 0.14cm, row sep = \rs]
			&\gate{U3(\theta_1,\theta_2,\theta_3)}&\ctrl{1}&\gate{R_Y(\theta_7)}&\targ{}&\gate{R_Y(\theta_9)}&\ctrl{1}&\gate{U3(\theta_{10},\theta_{11},\theta_{12})}&\\
			&\gate{U3(\theta_4,\theta_5,\theta_6)}&\targ{}&\gate{R_Y(\theta_8)}&\ctrl{-1}&&\targ{}&\gate{U3(\theta_{12},\theta_{13},\theta_{14})}&\\
		\end{quantikz}
	\label{F:SU4}}\\
	\subfigure[]{
		\begin{quantikz}[column sep = 0.14cm, row sep = \rs]
			&\gate{R_Y(\theta_1)}&\ctrl{1}&\gate{R_Y(\theta_3)}&\ctrl{1}&\gate{R_Y(\theta_5)}& \\
			&\gate{R_Y(\theta_2)}&\targ{}&\gate{R_Y(\theta_4)}&\targ{}&\gate{R_Y(\theta_6)}&\\
		\end{quantikz}
	\label{F:SO4}}
	\subfigure[]{
		\begin{quantikz}[column sep = 0.14cm, row sep = \rs]
			&\ctrl{1}&\octrl{1}& \\
			&\gate{R_Z(\theta_1)}&\gate{R_X(\theta_2)}&\\
		\end{quantikz}
	\label{F:Pooling}}
	\caption{Subunit circuits for the QCNN ansatz. (a) General $SU(4)$ unitary operator, where $U3(\theta_1,\theta_2,\theta_3) \eqp R_Z(\theta_2)HS^\dagger HR_Z(\theta_1)HSHR_Z(\theta_3)$. (b) General $SO(4)$ unitary operator. (c) Simple subunit used for pooling.}
	\label{F:unitaries}
\end{figure}

The subunits are unitary operators from the group $SU(4)$. While Figure \ref{F:SU4} shows the optimal general implementation \cite{SU4}, we adopt simpler configurations for efficiency. For convolutional layers, we use Figure \ref{F:SO4}---$SO(4)$ operators \cite{SO4}. For pooling layers, we use Figure \ref{F:Pooling}---controlled gates acting on qubit pairs \cite{QCNN}. Additionally, to further reduce parameters, subunits within each layer share identical values, ensuring the parameter count scales logarithmically with system size. This parametrization makes training feasible \cite{QCNN}.

\subsection{Quantum Function, Loss Function and Training Loop}

From the last qubit of the QCNN we obtain the output of the quantum neural network. For binary classification, we sample the probability of measuring the last qubit in state $\left|1\right>$. If this probability is higher than $0.5$, we assign $1$ to the predicted label, and $0$ otherwise. From a formal point of view, this probability is obtained by the following quantum function:
\begin{equation}\label{function}
	f(\mathbf{x},\boldsymbol{\theta}) = \frac{1-\left<Z_n\right>(\mathbf{x},\boldsymbol{\theta})}{2} ,
\end{equation}
where the subindex $n$ makes reference to the last qubit of the circuit, and
\begin{equation}
	\left<Z_n\right>(\mathbf{x},\boldsymbol{\theta}) = \left<\psi_\mathbf{x}\right|U(\boldsymbol{\theta})^\dagger Z_nU(\boldsymbol{\theta})\left|\psi_\mathbf{x}\right>.
\end{equation}

To fit this function to the training data, we need a loss function to minimize. The total loss is the mean loss over all training samples:
\begin{equation}\label{loss}
	\mathcal{L}(\boldsymbol{\theta}) = \frac{1}{|X|}\sum_X l(\mathbf{x},\boldsymbol{\theta}).
\end{equation}
A naive loss $l(\mathbf{x},\boldsymbol{\theta})$ for the data is the square error between the function output and the true label $y$:
\begin{equation}
	l(\mathbf{x},\boldsymbol{\theta}) = \left[f(\mathbf{x},\boldsymbol{\theta}) - y\right]^2.
\end{equation}
However, for classification tasks it is usual to use the cross-entropy loss \cite{Cross-Entropy}:
\begin{equation}\label{cross}
	l(\mathbf{x},\boldsymbol{\theta}) = - \left[y \log(f(\mathbf{x},\boldsymbol{\theta})) + (1 - y) \log(1 - f(\mathbf{x},\boldsymbol{\theta}))\right],
\end{equation}
which is minimum when the probability is equal to the binary label $y$.\\

To minimize this function we need to calculate its gradient, i.e., the derivative with respect each parameter. For this, different methods have been proposed, from finite differences and stochastic methods, to analytic methods using the parameter-shift rule\cite{Shift-rule}. In Section \ref{sec:Training} we show an example of implementation with the parameter-shift rule.\\

Once the derivatives have been obtained, the parameters are updated using a gradient-based method. The naive one is the gradient descend method, whose update rule at each time step $t$ is
\begin{equation}
	\theta_i^{(t)} = \theta_i^{(t-1)} - \alpha\frac{\partial \mathcal{L}}{\partial \theta_i}\left(\theta^{(t-1)}\right),
\end{equation}
where $\alpha$ is the learning rate, an hyperparameter that must be chosen case to case. Instead of using this naive approach, in our work we use \textit{Adam} method \cite{Adam}, which is usually used for classical and quantum neural networks with good performance. This method uses three additional hyperparameters, whose usual values are $\beta_1 = 0.9$, $\beta_2 = 0.999$ and $\epsilon = 10^{-8}$. The update rule for the parameters is given by the following equations:
\begin{equation}
	m_i^{(t)} = \beta_1\, m_i^{(t-1)} + (1-\beta_1)\,\frac{\partial \mathcal{L}}{\partial \theta_i}\big(\theta^{(t-1)}\big),
\end{equation}
\begin{equation}
	v_i^{(t)} = \beta_2\, v_i^{(t-1)} + (1-\beta_2)\,\left(\frac{\partial \mathcal{L}}{\partial \theta_i}\big(\theta^{(t-1)}\big)\right)^2,
\end{equation}
\begin{equation}
	\hat m_i^{(t)} = \frac{m_i^{(t)}}{1-\beta_1^t}, 
	\qquad
	\hat v_i^{(t)} = \frac{v_i^{(t)}}{1-\beta_2^t},
\end{equation}
\begin{equation}
	\theta_i^{(t)} = \theta_i^{(t-1)} - \alpha\frac{\hat{m}_i^{(t)}}{\sqrt{\hat{v}_i^{(t)}}+\epsilon},
\end{equation}
where the initial values $m^{(0)}$ and $v^{(0)}$ are set to $0$.

\section{QHE for Neural Networks}\label{sec:QHE-QNN}

This work investigates two scenarios: reverse delegated training and private inference. To do so, in this section we first describe the efficient decomposition of the QCNN ansatz into Clifford+$T$ gates, then introduce the CQC-QHE simulator used for our implementations.

\subsection{Decomposition into Clifford+$T$ gates}\label{sec:T_count}

To implement the QHE scheme to the QNN it is necessary to decompose the ansatz into the Clifford+$T$ set. In this section we show how our ansatz for the QCNN is decomposed and count the number of $T$ gates in the circuit, to prove that the QHE scheme can be implemented efficiently.\\

From Figure \ref{F:unitaries} we can see that all contributions to the $T$-gates count come from the rotation gates. Therefore, let us start counting the number of rotation gates in the ansatz, and then see how these gates can be decomposed. Indeed, all rotation gates can be reduced to $R_Z$ gates, since $R_X(\boldsymbol{\theta}) = HR_Z(\boldsymbol{\theta})H$, and $R_Y(\boldsymbol{\theta}) = SHR_Z(\boldsymbol{\theta})HS^\dagger = SHR_Z(\boldsymbol{\theta})HSZ$. Let us denote as $\nR_C$ and $\nR_P$ to the number of $R_Z$ provided by a single convolutional and pooling subunit, respectively. These numbers are bounded by the number of $R_Z$ gates in the most general $SU(4)$ operator of Figure \ref{F:SU4}, which is $15$ since each $U3$ gate comprises three rotation gates. Now, we need to count how many subunits there are in the ansatz. For a circuit with $n$ qubits, the number of effective qubits $n'$ at each layer is divided by two until there is one qubit remaining. Therefore, there are $\log_2(n)$ convolutional and pooling layers before fully compacting the information. Each of these layers has $n'$ convolutional subunits, except the last layer with $n'=2$ where there is a single convolutional subunit; and $n'/2$ pooling subunits. Using reverse layer indexing---where $i=1$ labels the final layer and $i=\log_2(n)$ labels the first layer---the effective qubit count per layer is $n' = 2^i$. Consequently, the total $R_Z$-gate count in the ansatz is:
\begin{equation}\label{R_A}
	\nR_A = \nR_C\left[1 + \sum_{i=2}^{\log_2(n)} 2^i\right] + \nR_P\sum_{i=1}^{\log_2(n)} \frac{2^i}{2} = \nR_C(2n - 3) + \nR_P(n - 1).
\end{equation}
Since the number of $R_Z$ gates scales linearly with the number of qubits, assuming that $R_Z$ can be decomposed efficiently, then the ansatz can be implemented in the QHE scheme.\\

To decompose a $R_Z$ gate, we could use the Solovay-Kitaev theorem \cite{Solovay-Kitaev}, which states that any single qubit gate $U$ can be implemented in Clifford+$T$ gates with $\mathcal{O}(\log^c(1/\varepsilon))$ gates, where $c$ is a small constant approximately equal to $2$, and $\varepsilon$ is the error of the approximation using a finite number of gates \cite{Nielsen,MA_review_modern_physics}:
\begin{equation}\label{Rz_error}
	\varepsilon \geq \Vert \bar{U}-U \Vert,
\end{equation}
where $\bar{U}$ is the approximated gate. However, there is a more efficient algorithm for decomposing a $R_Z$ into Clifford+$T$ gates \cite{Rz_decomposition}, which in the worst case needs a number $\nT_{R}$ of $T$ gates
\begin{equation}
	\nT_{R} = 4\log_2(1/\varepsilon)+O(\log(\log(1/\varepsilon)).
\end{equation}
This error $\varepsilon$ is due to the approximation of a single $R_Z$ gate. The total error of the algorithm, $\epsilon$, scales linearly with the number of gates that are decomposed approximately, given by \eqref{R_A}. Therefore:
\begin{equation}
	\varepsilon = \frac{\epsilon}{\nR_C(2n - 3) + \nR_P(n - 1)},
\end{equation}
so that the cost for a single $R_Z$ is expressed as
\begin{equation}
	\nT_{R} = 4\log_2\left(\frac{\nR_C(2n - 3) + \nR_P(n - 1)}{\epsilon}\right)+O\!\left(\log\left(\frac{\nR_C(2n - 3) + \nR_P(n - 1)}{\epsilon}\right)\right).
\end{equation}
Thus, we can conclude that the decomposition of the $R_Z$ gates only provides a logarithmic overhead, and thus the efficiency of the ansatz is held since the number of rotation gates scales linearly.\\

\begin{figure}[hbpt]
	\centering
	\begin{quantikz}[column sep = 0.14cm, row sep = \rs]
		&\ctrl{1}&\octrl{1}&\ghost{X} \\
		&\gate{R_Z(\theta_1)}&\gate{R_X(\theta_2)}&\ghost{R_Z\left(-\frac{\theta_2}{2}\right)}\\
	\end{quantikz}
	=
	\begin{quantikz}[column sep = 0.14cm, row sep = \rs]
		&\ctrl{1}&&\ctrl{1}&&\gate{X}&\ctrl{1}&&\ctrl{1}&&\gate{X}& \\
		&\targ{}&\gate{R_Z\left(-\frac{\theta_1}{2}\right)}&\targ{}&\gate{R_Z\left(\frac{\theta_1}{2}\right)}&\gate{H}&\targ{}&\gate{R_Z\left(-\frac{\theta_2}{2}\right)}&\targ{}&\gate{R_Z\left(\frac{\theta_2}{2}\right)}&\gate{H}&\\
	\end{quantikz}
	\caption{Decomposition of the controlled rotation gates of the pooling subunit in Figure \ref{F:Pooling} into $R_Z$ gates.}
	\label{F:Pooling_2}
\end{figure}
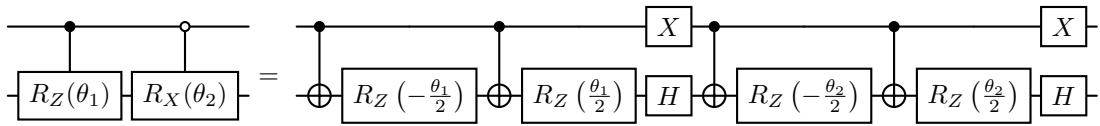

Finally, we specify concrete implementations for these subunits. For the  convolutional layer, we employ the $SO(4)$ circuit in Figure \ref{F:SO4}, which contains  $\nR_C = 6$ rotation gates per subunit. For the pooling layer we use the simple subunit in Figure \ref{F:Pooling}. This circuit has two controlled rotation gates. The controlled-$R_X$ gate can be transformed into a controlled-$R_Z$ gate with Hadamard gates, and since $R_Z(\boldsymbol{\theta}) = R_Z(\theta/2)XR_Z(-\theta/2)X$, this circuit can be decomposed into four $R_Z$ gates as shown in Figure \ref{F:Pooling_2}. Therefore, in our case $\nR_P = 4$, and the number of $T$ gates in our ansatz is
\begin{equation}
	\nT_A = \left(16n - 22\right)\left[ 4\log_2\left(\frac{16n - 22}{\epsilon}\right) 
	+ O\!\left(\log\left(\frac{16n - 22}{\epsilon}\right)\right) \right],
\end{equation}
which scales efficiently with the number of qubits and thus the QHE scheme can be implemented.

\subsection{QHE Simulation with CQC-QHE}

In the following sections we are going to show simulations results of two different scenarios using quantum homomorphic encryption. For this, we use CQC-QHE \cite{Homomorphic}, which is a python library we developed to simulate the QHE scheme in \cite{Liang}.\\

This package, based on Qiskit \cite{Qiskit}, takes a quantum circuit decomposed into the Clifford+$T$ set, and constructs a classical-quantum circuit that implements the homomorphic evaluation of each gate in SERVER's circuit, at the same time that simulates the key-updating process of CLIENT using classical and quantum operations. Moreover, we have now provided it with the ability to automatically compile circuits with arbitrary rotation gates into the Clifford+$T$ set, coupling it with the program \textit{gridsynth} \cite{Rz_decomposition}, which implements the efficient algorithm mentioned before.\\

Regarding memory efficiency, our simulator reuses a fixed two-qubit Bell register for all $T$-gate operations, yielding $\mathcal{O}(1)$ memory complexity independent of T-gate count. This enables simulation of circuits with arbitrarily many T gates. Nevertheless, due to the overhead for emulating classical operations with quantum gates, the time required to run the simulation is considerably increased, so that we may keep on small proof-of-concept examples in this work.

\section{Reverse Delegated Training}\label{sec:Training}

The first application of quantum homomorphic encryption in the field of quantum neural networks that we explore in this work is reverse delegated training. In contrast to a user with low quantum capabilities asking a server to train with both user's network and data, in this case a user wants to train a quantum neural network using its own quantum computer. However, it lacks the necessary data to do it, so there is another party that provides the data protecting it with homomorphic encryption. As an example, think of a laboratory that is doing research on QNN for diagnosis, and a hospital provides patient's data. Moreover, we explore the case where there is more than a single data provider, in a similar way of a federated training \cite{QNN-QHE-6}.\\

\begin{figure}[hbpt]
	\centering
	\makebox[0pt][c]{
	\includegraphics[width=0.7\linewidth]{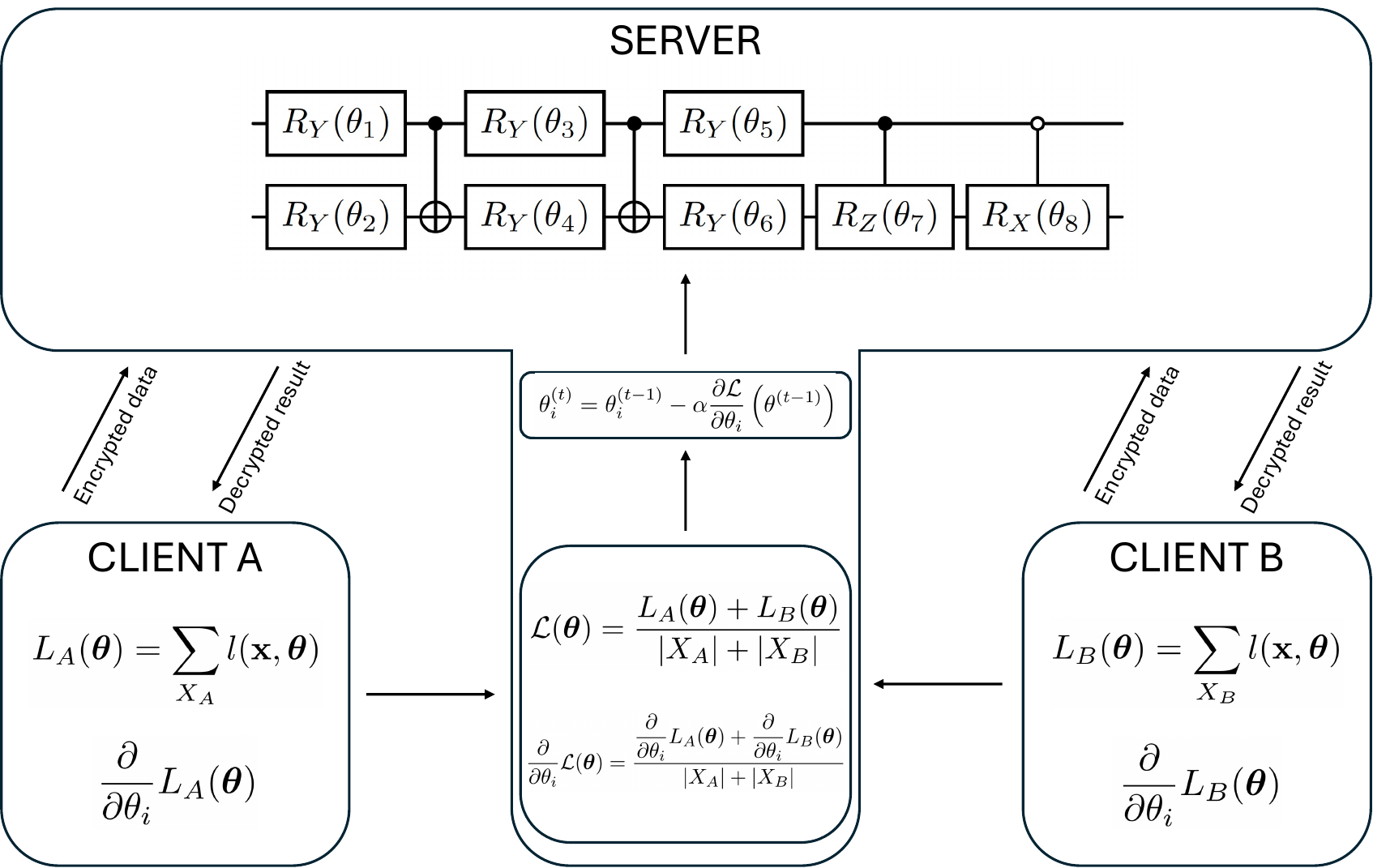}
	}
	\caption{Schematic of the reverse delegated training protocol. Multiple data providers (CLIENTs) supply encrypted datasets for the user (SERVER) to run the quantum neural network on them. After that, the data providers decrypt the results and obtain the gradients of the loss, which the user aggregates to train its quantum neural network.}
	\label{F:training}
\end{figure}

It should be noted that, while the data provider delivers a service to the neural network holder from a business standpoint, from a computational perspective the roles are reversed: within the QHE framework, the QNN holder functions as the SERVER, as it executes the algorithm, whereas the data provider acts as the CLIENT, supplying the encrypted data and subsequently decrypting the output \cite{Liang,Homomorphic}. For simplicity, let us suppose that we have two data providers, denoted as CLIENT A and CLIENT B. Each of these ones provides data from their training datasets $X_A$ and $X_B$, respectively. Therefore, we can express the loss function of the QNN in \eqref{loss} as
\begin{equation}
	\mathcal{L}(\boldsymbol{\theta})
	= \frac{1}{|X_A|+|X_B|} \left(\sum_{X_A} l(\mathbf{x},\boldsymbol{\theta}) + \sum_{X_B} l(\mathbf{x},\boldsymbol{\theta})\right) = \frac{L_A(\boldsymbol{\theta}) + L_B(\boldsymbol{\theta})}{|X_A|+|X_B|}.
\end{equation}
where $L_k(\boldsymbol{\theta})$ is the cumulative loss for the subset $X_k$. Since this expression has a separate term per CLIENT, then each CLIENT can calculate its loss with the decrypted results and pass it to SERVER for aggregating it. The same is true for calculating the derivatives for the training loop, since
\begin{equation}
	\frac{\partial}{\partial \theta_i}\mathcal{L}(\boldsymbol{\theta})
	= \frac{1}{|X_A|+|X_B|} \left( \frac{\partial}{\partial \theta_i} L_A(\boldsymbol{\theta}) + \frac{\partial}{\partial \theta_i} L_B(\boldsymbol{\theta}) \right).
\end{equation}
Therefore, SERVER can update its parameters after aggregating the individual derivatives. A scheme of the whole process is shown in Figure \ref{F:training}.\\

In our case we homomorphically train a simple QNN of two qubits using qubit encoding. The ansatz is composed of the general $SO(4)$ gate in Figure \ref{F:SO4} and the simple pooling layer of Figure \ref{F:Pooling}, so that there are in total $8$ parameters to train. Since we use the cross-entropy loss \eqref{cross}, the derivatives of each cumulative loss is
\begin{equation}\label{derivative}
	\frac{\partial}{\partial \theta_i}L_k(\boldsymbol{\theta}) = \sum_{X_k}\left[\frac{f(\mathbf{x},\boldsymbol{\theta}) - y}{f(\mathbf{x},\boldsymbol{\theta})[1 - f(\mathbf{x},\boldsymbol{\theta})]}
	\frac{\partial}{\partial \theta_i}f(\mathbf{x},\boldsymbol{\theta})\right].
\end{equation}
Since the quantum function $f(\mathbf{x},\boldsymbol{\theta})$ in \eqref{function} is an affine transformation of the expectation value of a Pauli operator, we can calculate the derivatives with the parameter-shift rule \cite{Shift-rule}. For the parameters in simple Pauli rotation gates we have:
\begin{equation}
	\frac{\partial}{\partial \theta_i} f(\mathbf{x},\boldsymbol{\theta})
	= \frac{1}{2}\left[ f\left(\mathbf{x},\boldsymbol{\theta} + \frac{\pi}{2} \boldsymbol{e}_i\right) - f\left(\mathbf{x},\boldsymbol{\theta} - \frac{\pi}{2} \boldsymbol{e}_i\right) \right],
\end{equation}
whereas for the parameters in controlled rotation gates we have \cite{Shift-rule-2}:
\begin{equation}
	\frac{\partial}{\partial \theta_i} f(\mathbf{x},\boldsymbol{\theta})
	= \frac{\sqrt{2}+1}{4\sqrt{2}} \left[ f\left(\mathbf{x},\boldsymbol{\theta} + \frac{\pi}{2} \boldsymbol{e}_i\right) - f\left(\mathbf{x},\boldsymbol{\theta} - \frac{\pi}{2} \boldsymbol{e}_i\right) \right]
	- \frac{\sqrt{2}-1}{4\sqrt{2}} \left[ f\left(\mathbf{x},\boldsymbol{\theta} + \frac{3\pi}{2} \boldsymbol{e}_i\right) - f\left(\mathbf{x},\boldsymbol{\theta} - \frac{3\pi}{2} \boldsymbol{e}_i\right) \right].
\end{equation}
Since there are six simple rotations and two controlled rotations, calculating the derivatives of the quantum function requires evaluating it $20$ times for each data sample $(\mathbf{x},y)$. An additional evaluation is required for the usual evaluation which involves \eqref{derivative}, to that in total $21$ evaluations are required. For the optimization loop we use the \textit{Adam} update rules, with a learning rate of $\alpha = 0.01$.\\

\begin{figure*}[hbpt]
	\centering
	\makebox[0pt][c]{
	\subfigure[]{\includegraphics[scale=0.5]{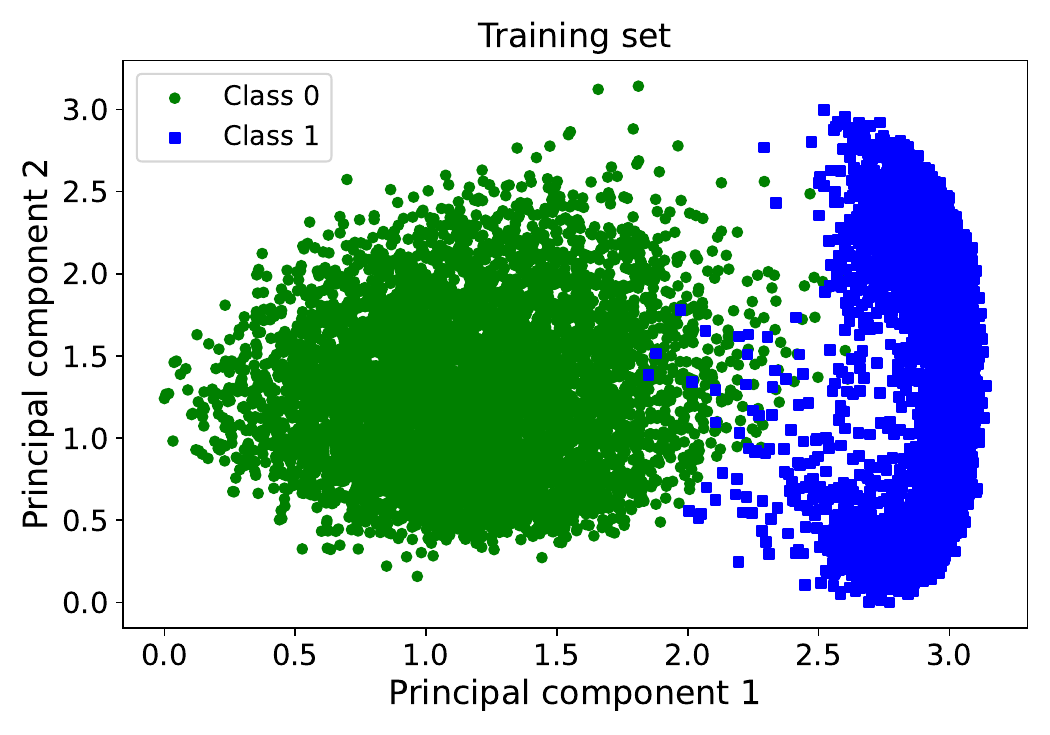}}
	\subfigure[]{\includegraphics[scale=0.5]{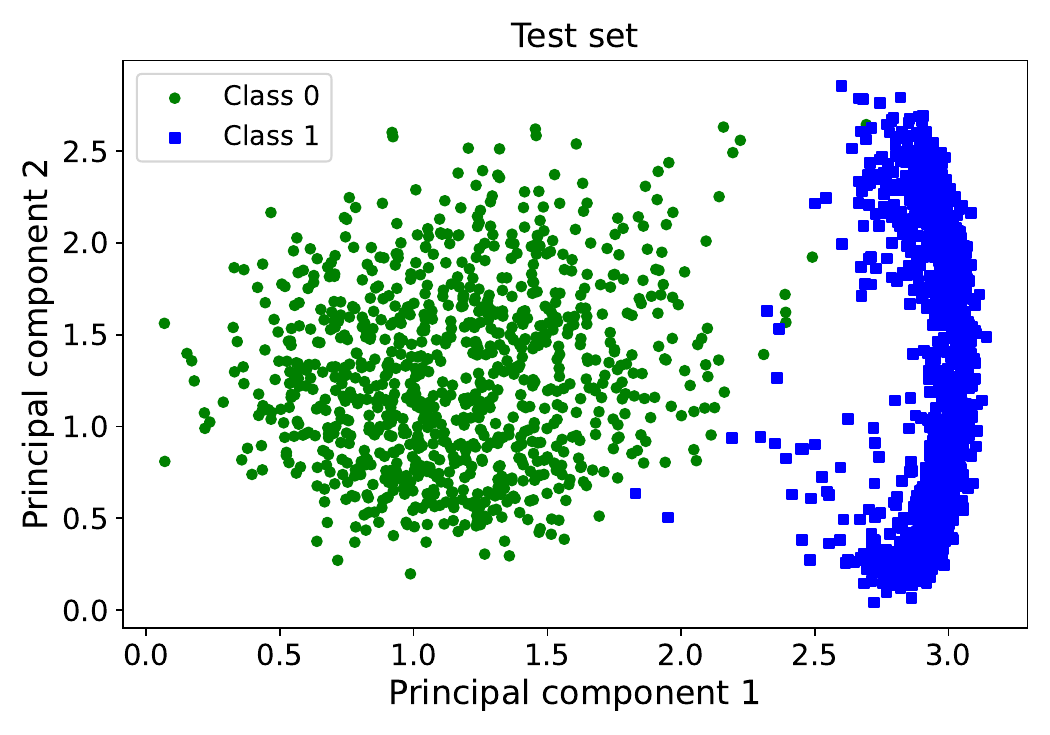}}
	}
	\caption{Two-dimensional representation of (a) the training and (b) the test datasets obtained via principal component analysis from MNIST digits 0 and 1.}
	\label{F:dataset}
\end{figure*}

For our simulation we take the handwritten digits $0$ and $1$ of the MNIST dataset from Scikit-learn \cite{Scikit-learn}. However, since our QNN is quite simple, more than actual images, what we actually feed to the circuit are 2D points obtained through principal component analysis (PCA) \cite{PCA}, normalized to be in $[0,\pi]^2$. The dataset has been divided with a relation $80/20$ for the training and test sets. Since we have reduced it to two-dimensional points, we can represent the datasets in charts, as shown in Figure \ref{F:dataset}. Moreover, to show an example of reverse federated learning where each data provider can indeed provide data from a particular class the provider is specialized on, in our simulation the dataset of CLIENT A contains all the images of the number $0$, and CLIENT B the number $1$.\\

\begin{figure}[hbpt]
	\centering
	\includegraphics[scale=0.6]{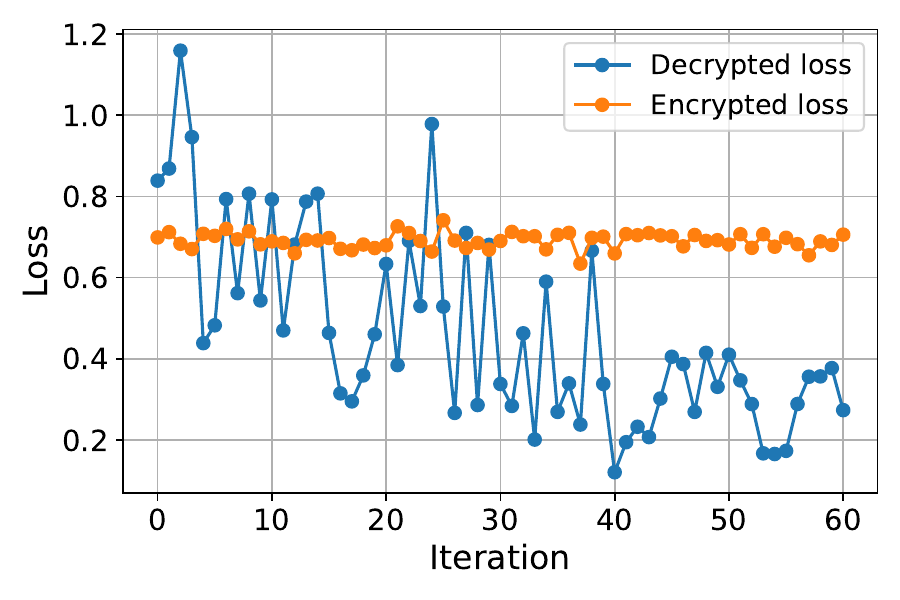}
	\caption{Training loss for each iteration of the reverse delegated training simulation. The decrypted loss (blue) decreases as training proceeds, while the encrypted loss (orange) remains constant around 0.693.}
	\label{F:loss}
\end{figure}

Following the training procedure in Ref. \cite{QCNN}, we use a small batch of data per iteration, rather than the entire training set. Indeed, to simplify the simulation, we only use two samples in the batch, so that it is composed of a sample of each CLIENT. At each iteration, each CLIENT obtains the encrypted results with $1024$ shots for each circuit evaluation, and decrypts them to calculate the loss and the gradient, which SERVER uses to minimize the loss. The loss curve of the training is shown in Figure \ref{F:loss}. There, we can see that the loss effectively decreases as the training proceeds. Moreover, we show what would happens if CLIENTs would not decrypt the results of the network. In that case, due to the QOTP encryption, the output probability would always be around $0.5$, not being exactly $0.5$ due to the sampling error with finite shots. Therefore, plugging it into \eqref{loss} and \eqref{cross}, the loss without decrypting would be always around $0.693$, and the gradient would be null, not being any optimization.\\

\begin{figure*}[hbpt]
	\centering
	\makebox[0pt][c]{
	\subfigure[]{\includegraphics[scale=0.5]{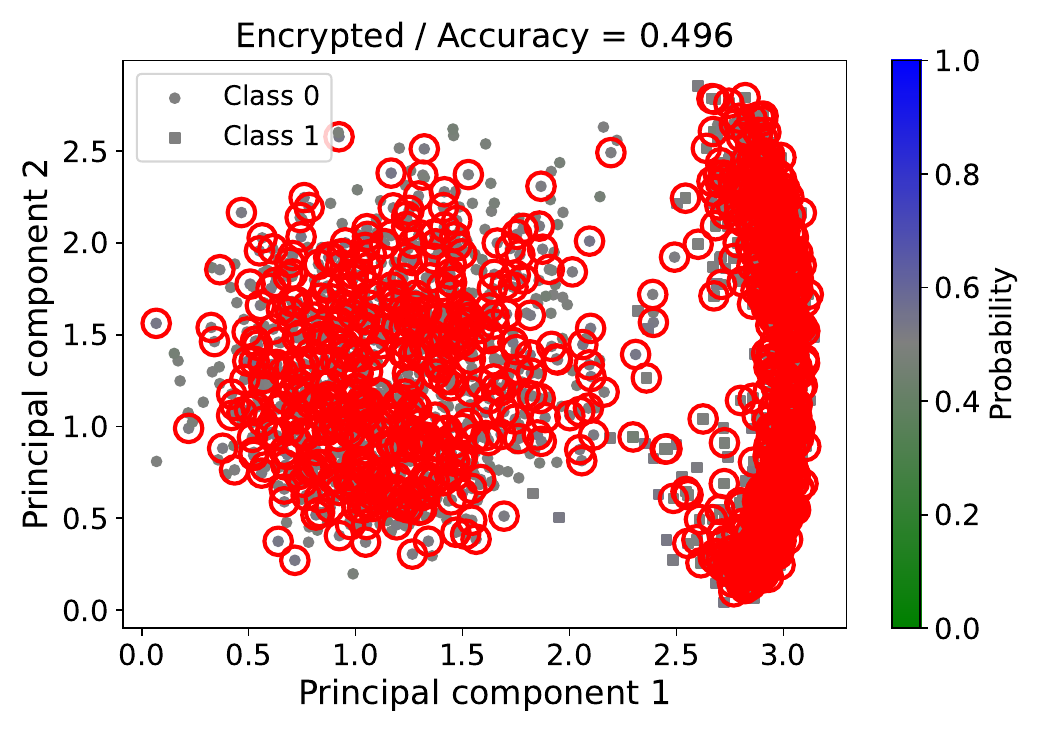}\label{F:test_e}}
	\subfigure[]{\includegraphics[scale=0.5]{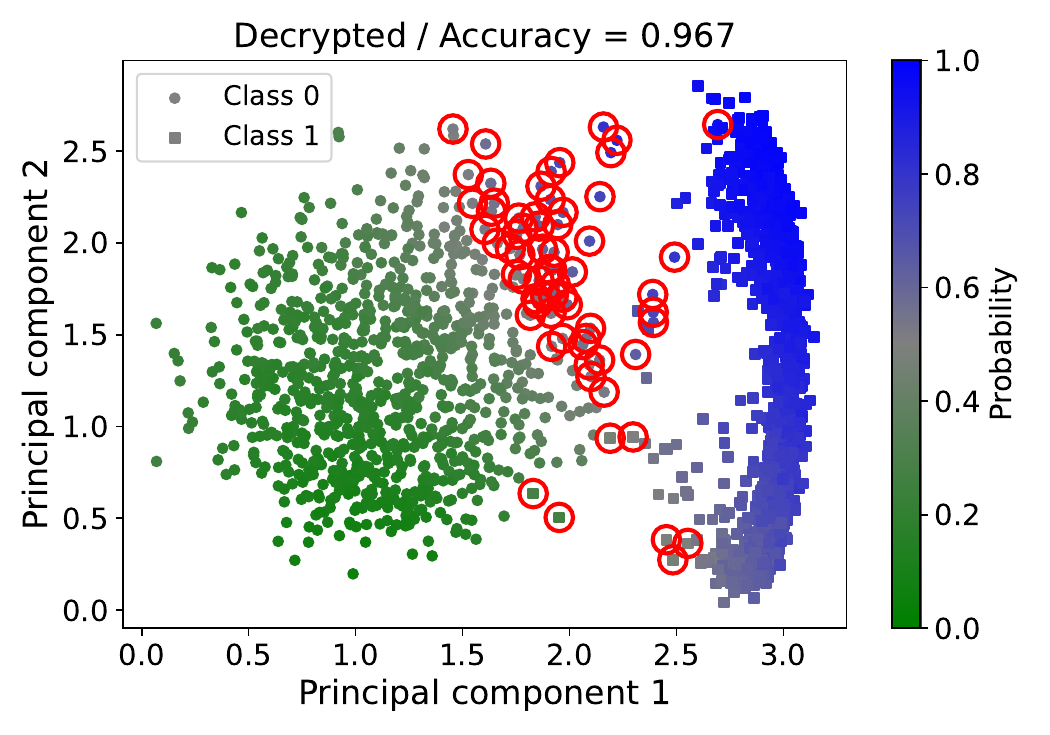}\label{F:test_d}}
	}
	\caption{Classification results on the test dataset after training the QCNN. (a) Encrypted predictions. (b) Decrypted predictions. Missclassified data points are surrounded by a red circle.}
	\label{F:test}
\end{figure*}

Once the network has been trained, we prove its accuracy on the test dataset. In Figure \ref{F:test_e} we show the predictions before decrypting, and in Figure \ref{F:test_d} after decrypting. Since the encrypted result produces a probability around $0.5$, which can be with the same chance above or below for the finite sampling effect, it is as if the QNN classified the data totally at random, and therefore an accuracy around $0.5$ is obtained. However, after decrypting, we can see that the network is able to classify with high accuracy, around $0.97$, so that it has been properly trained using the QHE scheme to protect the data.

\section{Private Inference}\label{sec:Inference}

The second application of the QHE scheme \cite{Liang} for QNN that we explore is private inference. In this case a user wants to make inference about some data using a private quantum neural network of a server. Note that in this case the terminology from QHE coincides with the business perspective, since the data holder is the actual client in this sense.\\

Since we are now interested in inference rather than training, we can take an already trained neural network for our simulation. Concretely, we have trained a network with eight qubits using Pennylane, which is a simulation framework specialized in quantum machine learning \cite{Pennylane}. Since we have more qubits than before, in this case we can input actual images from the MNIST dataset using amplitude encoding. However, since MNIST images are represented by $28 \times 28$ matrices, and with eight qubits we can only encode $2^8 = 256$ features, a simple classical preprocessing is done to resize the images to $16 \times 16$ matrices, which provide the $256$-dimensional feature vectors. The neural network trained in these conditions has an accuracy of $0.967$.\\

\begin{figure*}[hbpt]
	\centering
	\makebox[0pt][c]{\includegraphics[scale=0.5]{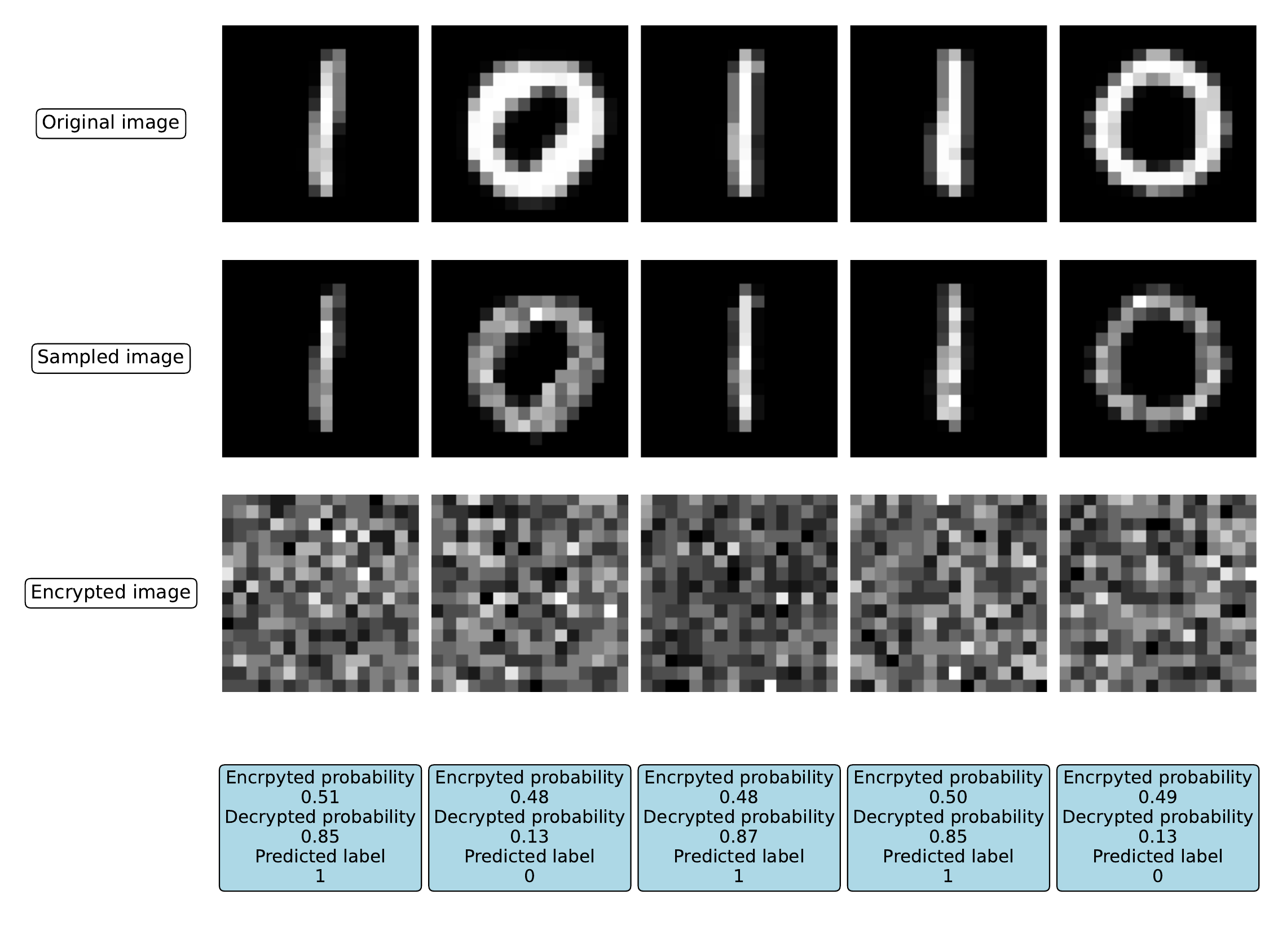}
	}
	\caption{Results of the private inference simulation using encrypted images from the MNIST dataset. The first row shows the original images, whereas the second and third rows show the sampled results from unencrypted and encrypted images, respectively. The last row shows the encrypted and decrypted output probabilities, demonstrating how the correct label, which corresponds to the particular handwritten digit, is obtained only after decryption.}
	\label{F:inference}
\end{figure*}

The Private Inference circuit is substantially larger than the previous Reverse Delegated Training circuit, making per-sample homomorphic simulation computationally expensive. Due to these constraints, we present detailed results for five representative test samples. Figure \ref{F:inference} displays these images (top row), pre-processed for the quantum circuit.\\

As discussed in Section \ref{sec:data}, amplitude encoding directly represents classical data as quantum amplitudes, allowing recovery through computational-basis measurements. \ref{F:inference} contrasts two scenarios: (row 2) Unencrypted state: 1024 measurements yield the original classical data, confirming direct recoverability. (row 3) Encrypted state (QOTP): The same measurements produce only uniform random  outcomes, demonstrating perfect data concealment. This illustrates how QOTP encryption completely masks the underlying data.\\

Finally, let us see the results after running the QNN, which are shown in the last row of Figure \ref{F:inference}. As expected, the probability sampled from the encrypted result is around $0.5$ in all cases. When the result is decrypted, then the true probability is obtained, and we can use it to predict the labels, which coincide with the true labels. Thus, a user can effectively take advantage of QHE to make inference on a server's network while protecting its data.

\subsection{Privacy of Server's Circuit}

So far, we have seen that CLIENT is able to make inference about its data using SERVER'S quantum neural network. However, for this QHE scheme to be useful, it is necessary that CLIENT does not obtain the circuit of the trained ansatz during the decryption. Otherwise, CLIENT would run the QNN by itself after the first homomorphic shot and would not need SERVER anymore, or could send such information to competitors, so that SERVER would loss the business.\\

As for the QHE scheme we have used \cite{Liang}, we have seen that CLIENT has to apply the decrypting functions corresponding to the gates of the circuit. Indeed, even if the deterministic functions corresponding to consecutive Clifford gates were composed into a single function, it would still very easy for CLIENT to decompose it \cite{Homomorphic}.  Therefore, CLIENT could use the key-updating functions to reconstruct the sequence of gates in the circuit of SERVER. Nevertheless, since for Pauli gates the key is not updated when they are applied, CLIENT is unable to know whether there are Pauli gates in the circuit, and where they would be located. Therefore, we can say that CLIENT obtains all the information of the ansatz except the presence of the Pauli gates $X$ and $Z$ in the quantum circuit.\\

In the ansatz we are using for the QNN, the contribution of Pauli gates mostly comes from decomposing the rotation gates into Clifford+$T$ gates. Therefore, the effect of lacking knowledge about Pauli gates may be equivalent to CLIENT not been able to obtain the true weights of the trained network. For this reason, in this section we explore how this lack of knowledge affects to the QNN that CLIENT can reconstruct from the decryption functions.\\

\begin{figure*}[hbpt]
	\centering
	\subfigure[]{\includegraphics[scale=0.5]{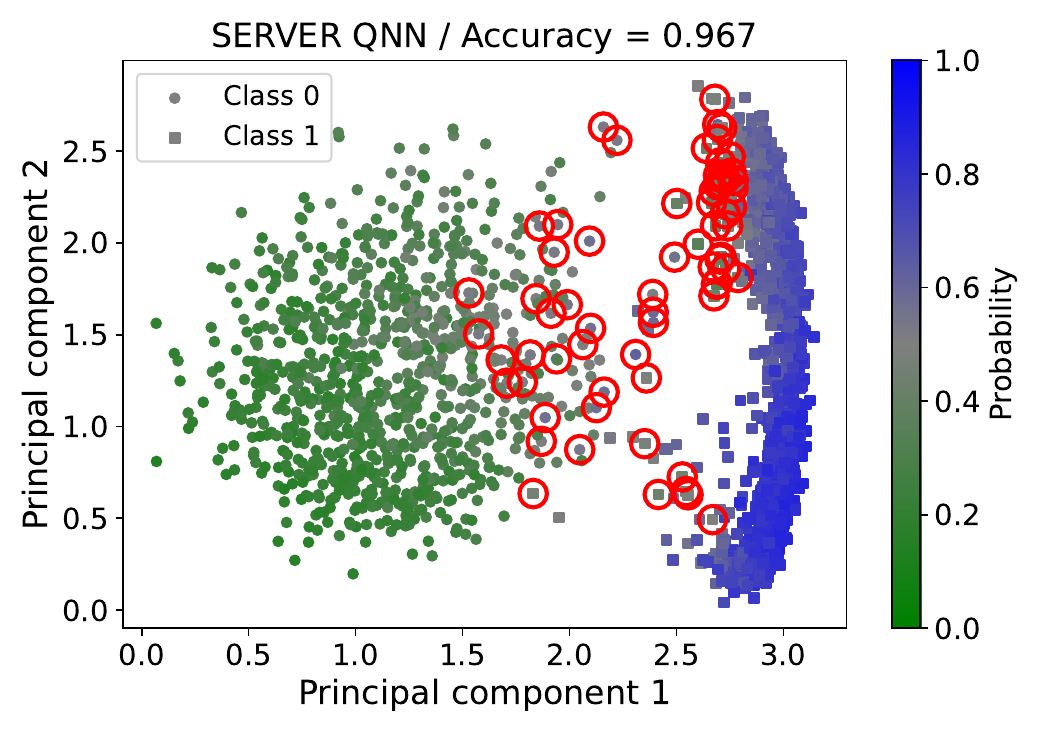}\label{F:pauli_1}}
	\subfigure[]{\includegraphics[scale=0.5]{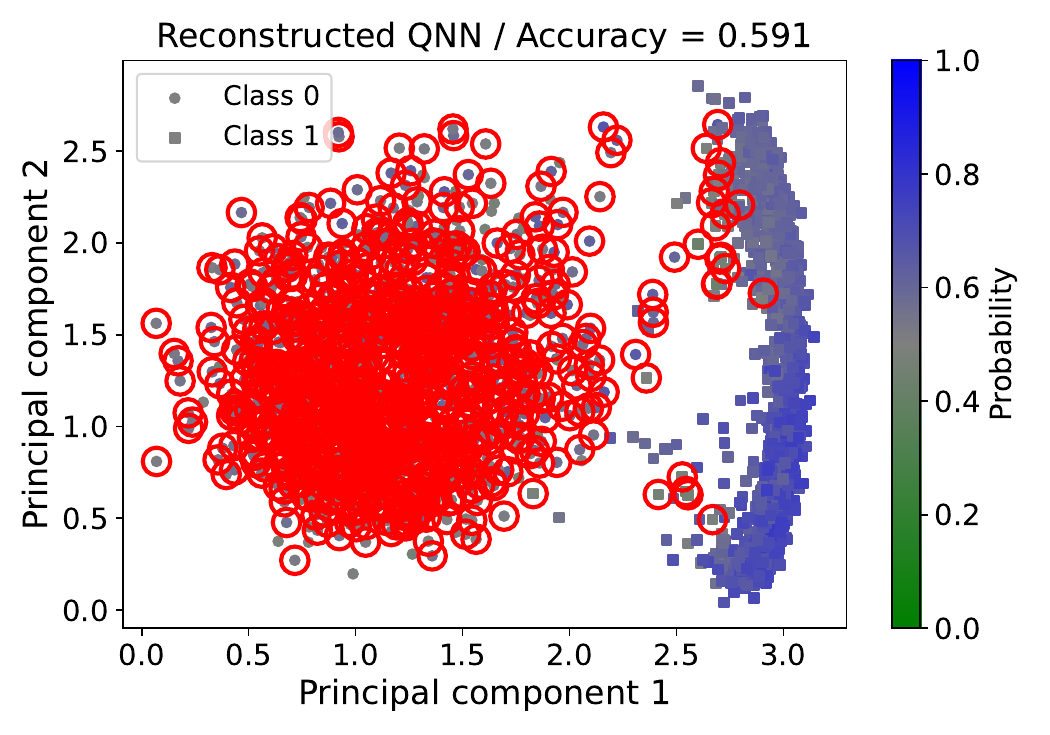}\label{F:pauli_2}}
	\caption{Analysis of SERVER's circuit privacy. (a) Classification performance of the original QCNN on the test set. (b) Performance of the reconstructed circuit after removing Pauli gates on the test set, showing significant accuracy degradation and thus preservation of model confidentiality. Missclassified data points are surrounded by a red circle.}
	\label{F:...}
\end{figure*}

Following with the QNN of eight qubits used before, let us first see graphically its accuracy on the test set representing with a chart the predicted probabilities over the equivalent 2D data obtained by PCA. In Figure \ref{F:pauli_1} we can see that effectively this QNN is able to properly classify most of the data. However, when we remove the Pauli gates from the circuit, the accuracy drops to $0.591$. Indeed, in Figure \ref{F:pauli_2} we observe that it seems that the reconstructed network is classifying each data sample as the number $1$, so that it is not able to distinguish between both classes.\\

Since the training process is stochastic (starting from randomly initialized weights), each training run produces different final weights, yet achieving similar accuracy across runs. The results presented above represent just one example of this variability.  However, the blindness of the network to the encrypted data could vary across different trained instances. In Table \ref{tab:pauli} we show the accuracy obtained after training and after removing the Pauli gates for $10$ instances of QNN with eight qubits, using both amplitude and qubit encoding. In all cases we see that a high accuracy is reached after training. However, not all the instances have the same accuracy drop when removing the Pauli gates. Note that, since a network with null accuracy would assign the opposite label to each data sample, CLIENT could reconstruct a functional network just inverting the output. Therefore, the network is more blind to the data when the accuracy is nearer to $0.5$. Thus, to quantify the effect in the accuracy, we actually look at the distance from $0.5$ in absolute value. The nearer this value is to $0$, the more private the network is. In most cases this value is under $0.2$, so that the network can be held private. However, there are some instances where the drop in accuracy is very small, so that CLIENT could still classify a great proportion of data with the reconstructed network.\\

\begin{table}[htbp]
	\centering
	\caption{Accuracy of trained quantum convolutional neural networks of SERVER, and of the reconstructed network without Pauli gates, for ten independent training instances, using both amplitude and qubit encodings. The column $|\text{Non-Pauli}-0.5|$ quantifies the distance from random-guess accuracy, illustrating how the absence of Pauli gates blinds the reconstructed network and preserves model privacy in most of the cases.}
	\begin{tabular}{c|ccc|ccc}
		\hline
		\multirow{2}[4]{*}{Index} & \multicolumn{3}{c|}{Amplitude Encoding} & \multicolumn{3}{c}{Qubit Encoding} \\
		\cline{2-7}      & SERVER & Non-Pauli & $|\text{Non-Pauli}-0.5|$ & SERVER & Non-Pauli & \multicolumn{1}{c|}{$|\text{Non-Pauli}-0.5|$} \\
		\hline
		1     & 0.967 & 0.788 & 0.288 & 0.991 & 0.440 & 0.060 \\
		2     & 0.986 & 0.572 & 0.072 & 0.988 & 0.486 & 0.015 \\
		3     & 0.983 & 0.885 & 0.385 & 0.994 & 0.446 & 0.054 \\
		4     & 0.992 & 0.390 & 0.111 & 0.989 & 0.256 & 0.244 \\
		5     & 0.978 & 0.886 & 0.386 & 0.983 & 0.536 & 0.036 \\
		6     & 0.982 & 0.322 & 0.179 & 0.990 & 0.170 & 0.330 \\
		7     & 0.984 & 0.579 & 0.079 & 0.972 & 0.598 & 0.098 \\
		8     & 0.973 & 0.887 & 0.387 & 0.985 & 0.535 & 0.035 \\
		9     & 0.989 & 0.856 & 0.356 & 0.990 & 0.448 & 0.053 \\
		10    & 0.984 & 0.440 & 0.060 & 0.983 & 0.185 & 0.316 \\
		\hline
	\end{tabular}
	\label{tab:pauli}
\end{table}

Although algorithmic privacy guarantees cannot be provided for all post-training weight configurations, SERVER can employ a practical defense: repeatedly retrain the model, evaluate accuracy across instances, and select the configuration with degraded accuracy (which indicates maximum privacy protection). Our preliminary results demonstrate that this approach consistently succeeds---revealing an important privacy-accuracy trade-off. Moreover, even if the reconstructed network retains a high accuracy around $0.8$ as in our examples, depending on the use case this can also be dreadful for the network. For example, think of medical diagnosis, where a really high accuracy near to $0.99$ is desirable to avoid false diagnostics. Therefore, despite the high accuracy, CLIENT may not be so sure about using the reconstructed network, so that SERVER do not lose its business.

\section{Conclusions}\label{sec:Conclusions}

We have studied the possibility of using quantum homomorphic encryption for protecting sensitive data in the context of quantum machine learning. In particular, we have focused on the use of the scheme developed in \cite{Liang}, which in contrast to other approaches previously used, is perfectly secure, at the same time that is non-interactive, $\mathcal{F}$-homomorphic and quasi-compact.\\

As an interesting example, we have focused on quantum convolutional neural networks. These networks not only have absence of barren plateaus, but also have simple circuits suitable for the NISQ era. Therefore, as we have proved, they can be decomposed efficiently into Clifford+$T$ gates, so that they can be implemented within a QHE scheme, even if it requires a further quantum error correction.\\

To simulate different applications of the QHE scheme for quantum machine learning, we have used the python library CQC-QHE, which is a library able to construct classical-quantum circuits for homomorphic simulation, and we have coupled it with a program able to automatically compile the ansatz circuits into Clifford+$T$ gates. Moreover, these simulations constitute the first simulations of this QHE scheme \cite{Liang} where the actual information is protected in the initial state, in contrast to other proof-of-concept simulations were the actual information was in the algorithm's circuit rather than in the initial state. Therefore, this proves that this QHE scheme is effectively useful for cases with real applications.\\

We have explored two applications of the scheme for QNN. The first one is reverse delegated training, where a user trains its network using its own quantum computer, and a data provider supplies the encrypted data. Moreover, we have shown how this scenario can be expanded to a federated learning, with several different data providers working at the same time. In our simulation, we have managed to train a simple 2-qubit neural network with 2D data supplied by two different data providers within the QHE scheme.\\

The second application we explored is private inference, where a user makes inference on its data using a server's private quantum neural network in remote. Using a QCNN with eight qubits and handwritten digits images from the MNIST dataset, we have showed that the user can effectively protect the image and obtain the correct prediction after decrypting the output of the network. Furthermore, we have also analyzed whether the server is able to keep secret the weights of the network from the client, observing that depending on the weights obtained after training, there is a high chance that the information is kept secret.\\

Although there are some instances where trained networks are not kept perfectly private from the client, we would like to emphasize that we have only used networks with $24$ parameters so far. However, for more complex networks, with more qubits and parameters, the number of Pauli gates increases when decomposing the rotation gates. Therefore, we expect a bigger effect on the accuracy when removing Pauli gates, so that the privacy of the network is ensured. Furthermore, note than in this work we have used \textit{gridsynth} for the decomposition. However, we could look for other decomposition methods that, although not being optimal, may provide more Pauli gates so that the effect is bigger when removing them.\\

In the future, it would be interesting to study the implementation of QHE for other QNN architectures, thus widening the range of applications. Moreover, we must also go deeper in the privacy of the PQC. For example, we could develop special QNN architectures with Pauli gates places cleverly so that the accuracy effectively drops when they are removed, or develop new QHE schemes focused on improving the current privacy issues of current schemes.

\section{Data Availability Statement}\label{Data}

The codes for the simulation results are available on \url{https://github.com/OrtegaSA/qnn-cqc}.

The new version of the library CQC-QHE is available on \url{https://github.com/OrtegaSA/CQC-QHE-repo}.

\section*{Acknowledgments}

We acknowledge the support from the Spanish MINECO grants MINECO/FEDER Projects, PID2021-122547NB-I00 FIS2021, the “MADQuantum-CM” project funded by Comunidad de Madrid and by the Recovery, Transformation, and Resilience Plan – Funded by the European Union - NextGenerationEU and Ministry of Economic Affairs Quantum ENIA project. This work has also been financially supported by the Ministry for Digital Transformation and of Civil Service of the Spanish Government through the QUANTUM ENIA project call – Quantum Spain project, and by the European Union through the Recovery, Transformation and Resilience Plan – NextGenerationEU within the framework of the Digital Spain 2026 Agenda. M. A. M.-D. has been partially supported by the U.S.Army Research Office through Grant No. W911NF-14-1-0103. S.A.O. acknowledges support from Universidad Complutense de Madrid - Banco Santander through Grant No. CT58/21-CT59/21.

\bibliography{MiBiblio}

\begin{thebibliography}{10}

\bibitem{Deep_learning}
Y.~LeCun, Y.~Bengio, and G.~Hinton.
\newblock {Deep Learning.}
\newblock {\em Nature}, 521:436--444, 2015.

\bibitem{QASUML}
S.~Lloyd, M.~Mohseni, and P.~Rebentrost.
\newblock {Quantum Algorithms for Supervised and Unsupervised Machine
  Learning.}
\newblock {\em arXiv:1307.0411}, 2013.

\bibitem{Paparo3}
G.~D. Paparo, V.~Dunjko, A.~Makmal, M.~A. Martin-Delgado, and H.~J. Briegel.
\newblock {Quantum Speedup for Active Learning Agents}.
\newblock {\em Physical Review X}, 4:031002, 2014.

\bibitem{Power_QNN}
A.~Abbas, D.~Sutter, C.~Zoufal, A.~Lucchi, A.~Figalli, and S.~Woerner.
\newblock {The Power of Quantum Neural Networks.}
\newblock {\em Nature Computational Science}, 1:403--409, 2021.

\bibitem{QML}
J.~Biamonte, P.~Wittek, N.~Pancotti, P.~Rebentrost, N.~Wiebe, and S.~Lloyd.
\newblock {Quantum Machine Learning}.
\newblock {\em Nature}, 549:195--202, 2017.

\bibitem{QML-2}
M.~Schuld, I.~Sinayskiy, and F.~Petruccione.
\newblock {An Introduction to Quantum Machine Learning.}
\newblock {\em Contemporary Physics}, 56:172--185, 2015.

\bibitem{QNN}
Y.~Kwak, W.~J. Yun, S.~Jung, and J.~Kim.
\newblock {Quantum Neural Networks: Concepts, Applications, and Challenges}.
\newblock In {\em 2021 Twelfth International Conference on Ubiquitous and
  Future Networks (ICUFN). IEEE}, pages 413--416, 2021.

\bibitem{Preskil}
J.~Preskill.
\newblock {Quantum Computing in the NISQ Era and Beyond}.
\newblock {\em Quantum}, 2:79, 2018.

\bibitem{NISQA}
K.~Bharti~\textit{et al}.
\newblock {Noisy Intermediate-Scale Quantum Algorithms.}
\newblock {\em Reviews of Modern Physics}, 94:015004, 2022.

\bibitem{VQA}
M.~Cerezo~\textit{et al}.
\newblock {Variational Quantum Algorithms.}
\newblock {\em Nature Reviews Physics}, 3:625--644, 2021.

\bibitem{PQCML}
M.~Benedetti, E.~Lloyd, S.~Sack, and M.~Fiorentini.
\newblock {Parameterized Quantum Circuits as Machine Learning Models}.
\newblock {\em Quantum Science and Technology}, 4:043001, 2019.

\bibitem{QFHE_def}
M.~Liang.
\newblock {Symmetric Quantum Fully Homomorphic Encryption with Perfect
  Security}.
\newblock {\em Quantum Information Processing}, 12:3675--3687, 2013.

\bibitem{T_interactions}
M.~Liang.
\newblock {Quantum Fully Homomorphic Encryption Scheme Based on Universal
  Quantum Circuit}.
\newblock {\em Quantum Information Processing}, 14:2749--2759, 2015.

\bibitem{Broadbent}
A.~Broadbent and S.~Jeffery.
\newblock {Quantum Homomorphic Encryption for Circuits of Low T-Gate
  Complexity}.
\newblock In {\em Advances in Cryptology -- CRYPTO 2015}, pages 609--629, 2015.

\bibitem{No_go_result}
L.~Yu, C.~A. Pérez-Delgado, and J.~F. Fitzsimons.
\newblock {Limitations on Information Theoretically Secure Quantum Homomorphic
  Encryption}.
\newblock {\em Physical Review A}, 90:050303, 2014.

\bibitem{Enhanced_no_go}
C.~Y. Lai and K.~M. Chung.
\newblock {On Statistically-Secure Quantum Homomorphic Encryption}.
\newblock {\em Quantum Information and Computation}, 18:0785--0794, 2018.

\bibitem{Liang}
M.~Liang.
\newblock {Teleportation-Based Quantum Homomorphic Encryption Scheme with
  Quasi-Compactness and Perfect Security.}
\newblock {\em Quantum Information Processing}, 19:28, 2020.

\bibitem{QNN-QHE-1}
W.~Sun, Y.~Chang, D.~Wang, S.~Zhang, and L.~Yan.
\newblock {Delegated Quantum Neural Networks for Encrypted Data.}
\newblock {\em Physica Scripta}, 99:055102, 2024.

\bibitem{QNN-QHE-2}
L.~Zeng, Y.~Chang, X.~Zhang, W.~Xue, S.~Zhang, L.~Yan, and Z.~Gou.
\newblock {Distributed Machine Learning Based on Quantum Cloud with Quantum
  Homomorphic Encryption.}
\newblock {\em Future Generation Computer Systems}, 175:108053, 2025.

\bibitem{QNN-QHE-3}
S.~Dutta~\textit{et al}.
\newblock {Federated Learning with Quantum Computing and Fully Homomorphic
  Encryption: A Novel Computing Paradigm Shift in Privacy-Preserving ML}.
\newblock {\em arXiv:2409.11430}, 2024.

\bibitem{QNN-QHE-5}
C.~Fang and Y.~Chang.
\newblock {Quantum Neural Network with Privacy Protection of Input Data and
  Training Parameters}.
\newblock {\em Physica Scripta}, 99:035111, 2024.

\bibitem{QNN-QHE-6}
W.~Li and D.~L. Deng.
\newblock {Quantum Delegated and Federated Learning Via Quantum Homomorphic
  Encryption}.
\newblock {\em Research Directions: Quantum Technologies}, 3:e3, 2025.

\bibitem{QNN-QHE-7}
Q.~Li, J.~Quan, J.~Shi, S.~Zhang, and X.~Li.
\newblock {Secure Delegated Variational Quantum Algorithms}.
\newblock {\em IEEE Transactions on Computer-Aided Design of Integrated
  Circuits and Systems}, 43:3129--3142, 2024.

\bibitem{CNN}
Y.~LeCun, B.~Boser, J.~S. Denker, D.~Henderson, R.~E. Howard, W.~Hubbard, and
  L.~D. Jackel.
\newblock {Backpropagation Applied to Handwritten Zip Code Recognition.}
\newblock {\em Neural Computation}, 1:541--551, 1989.

\bibitem{Homomorphic}
S.~A. Ortega, P.~Fernández, and M.~A. Martin-Delgado.
\newblock {Implementing Semiclassical Szegedy Walks in Classical-Quantum
  Circuits for Homomorphic Encryption}.
\newblock {\em Journal of Physics: Complexity}, 6:025010, 2025.

\bibitem{QHE_grover}
C.~Gong, J.~Du, Z.~Dong, Z.~Guo, A.~Gani, L.~Zhao, and H.~Qi.
\newblock {Grover Algorithm-Based Quantum Homomorphic Encryption Ciphertext
  Retrieval Scheme in Quantum Cloud Computing.}
\newblock {\em Quantum Information Processing}, 19:1--17, 2020.

\bibitem{Pablo_Grover}
P.~Fernández and M.~A. Martin-Delgado.
\newblock {Implementing the Grover Algorithm in Homomorphic Encryption
  Schemes}.
\newblock {\em Physical Review Research}, 6:043109, 2024.

\bibitem{Grover}
L.~K. Grover.
\newblock {A Fast Quantum Mechanical Algorithm for Database Search}.
\newblock In {\em Proceedings of the 28th Annual ACM Symposium on Theory of
  Computing}, pages 212--219, 1996.

\bibitem{Szegedy}
M.~Szegedy.
\newblock {Quantum Speed-up of Markov Chain Based Algorithms}.
\newblock In {\em 45th Annual IEEE Symposium on Foundations of Computer
  Science}, pages 32--41, 2004.

\bibitem{One_time_pad}
P.~O. Boykin and V.~Roychowdhury.
\newblock {Optimal Encryption of Quantum Bits.}
\newblock {\em Physical Review A}, 67:042317, 2003.

\bibitem{Nielsen}
M.~A. Nielsen and I.~L. Chuang.
\newblock {\em {Quantum Computation and Quantum Information}}.
\newblock Cambridge University Press, 2010.

\bibitem{Error_correction}
A.~Berthiaume, D.~Deutsch, and R.~Jozsa.
\newblock {The Stabilisation of Quantum Computations}.
\newblock In {\em Proceedings Workshop on Physics and Computation. PhysComp'94,
  IEEE}, pages 60--62, 1994.

\bibitem{T-complexity}
C.~Yuan and M.~Carbin.
\newblock {The T-Complexity Costs of Error Correction for Control Flow in
  Quantum Computation}.
\newblock {\em Proceedings of the ACM on Programming Languages}, 8:492--517,
  2024.

\bibitem{TFermion}
P.~A.~M. Casares, R.~Campos, and M.~A. Martin-Delgado.
\newblock {TFermion: A Non-Clifford Gate Cost Assessment Library of Quantum
  Phase Estimation Algorithms for Quantum Chemistry}.
\newblock {\em Quantum}, 6:768, 2022.

\bibitem{QSVDD}
H.~Oh and D.~K. Park.
\newblock {Quantum Support Vector Data Description for Anomaly Detection.}
\newblock {\em Machine Learning: Science and Technology}, 5:035052, 2024.

\bibitem{Quantum_sensor}
M.~Cerezo, G.~Verdon, H.~Y. Huang, L.~Cincio, and P.~J. Coles.
\newblock {Challenges and Opportunities in Quantum Machine Learning.}
\newblock {\em Nature Computational Science}, 2:567--576, 2022.

\bibitem{Schuld_feature_space}
M.~Schuld and N.~Killoran.
\newblock {Quantum Machine Learning in Feature Hilbert Spaces.}
\newblock {\em Physical Review Letters}, 122:040504, 2019.

\bibitem{QML_book}
P.~A.~M. Dirac.
\newblock {\em {Supervised Learning with Quantum Computers}}.
\newblock Springer: Quantum Science and Technology, 2018.

\bibitem{QCNN}
T.~Hur, L.~Kim, and D.~K. Park.
\newblock {Quantum Convolutional Neural Network for Classical Data
  Classification.}
\newblock {\em Quantum Machine Intelligence}, 4:3, 2022.

\bibitem{Initializer}
R.~Iten, R.~Colbeck, I.~Kukuljan, J.~Home, and M~Christandl.
\newblock {Quantum Circuits for Isometries}.
\newblock {\em Physical Review A}, 93:032318, 2016.

\bibitem{ZZ}
V.~Havlíček, A.~D. Córcoles, K.~Temme, A.~W. Harrow, A.~Kandala, J.~M. Chow,
  and J.~M. Gambetta.
\newblock {Supervised Learning with Quantum-Enhanced Feature Spaces.}
\newblock {\em Nature}, 567:209--212, 2019.

\bibitem{No-barren}
A.~Pesah, M.~Cerezo, S.~Wang, T.~Volkoff, A.~T. Sornborger, and P.~J. Coles.
\newblock {Absence of Barren Plateaus in Quantum Convolutional Neural
  Networks.}
\newblock {\em Physical Review X}, 11:041011, 2021.

\bibitem{SU4}
F.~Vatan and C.~Williams.
\newblock {Optimal Quantum Circuits for General Two-Qubit Gates.}
\newblock {\em Physical Review A}, 69:032315, 2004.

\bibitem{SO4}
H.~R. Wei and Y.~M. Di.
\newblock {Decomposition of Orthogonal Matrix and Synthesis of Two-Qubit and
  Three-Qubit Orthogonal Gates.}
\newblock {\em arXiv:1203.0722}, 2012.

\bibitem{Cross-Entropy}
A.~Mao, M.~Mohri, and Y.~Zhong.
\newblock {Cross-Entropy Loss Functions: Theoretical Analysis and
  Applications.}
\newblock In {\em International Conference on Machine Learning}, pages
  23803--23828, 2023.

\bibitem{Shift-rule}
M.~Schuld, V.~Bergholm, C.~Gogolin, J.~Izaac, and N.~Killoran.
\newblock {Evaluating Analytic Gradients on Quantum Hardware.}
\newblock {\em Physical Review A}, 99:032331, 2019.

\bibitem{Adam}
D.~P. Kingma and J.~Ba.
\newblock {Adam: A Method for Stochastic Optimization.}
\newblock {\em arXiv:1412.6980}, 2014.

\bibitem{Solovay-Kitaev}
A.~Y. Kitaev.
\newblock {Quantum Computations: Algorithms and Error Correction}.
\newblock {\em Russian Mathematical Surveys}, 52:1191, 1997.

\bibitem{MA_review_modern_physics}
A.~Galindo and M.~A. Martin-Delgado.
\newblock {Information and Computation: Classical and Quantum Aspects}.
\newblock {\em Reviews of Modern Physics}, 74:347, 2002.

\bibitem{Rz_decomposition}
N.~J. Ross and P.~Selinger.
\newblock {Optimal Ancilla-Free Clifford+T Approximation of Z-Rotations.}
\newblock {\em Quantum Information and Computation}, 16:901--953, 2016.

\bibitem{Qiskit}
A.~Javadi-Abhari~\textit{et al}.
\newblock {Quantum Computing with Qiskit}.
\newblock {\em arXiv:2405.08810}, 2024.

\bibitem{Shift-rule-2}
G.~L.~R. Anselmetti, D.~Wierichs, C.~Gogolin, and R.~M Parrish.
\newblock {Local, Expressive, Quantum-Number-Preserving VQE Ansätze for
  Fermionic Systems.}
\newblock {\em New Journal of Physics}, 23:113010, 2021.

\bibitem{Scikit-learn}
F.~Pedregosa~\textit{et al}.
\newblock {Scikit-Learn: Machine Learning in Python.}
\newblock {\em Journal of Machine Learning Research}, 12:2825--2830, 2011.

\bibitem{PCA}
I.~T. Jolliffe.
\newblock {\em {Principal Component Analysis. Springer Series in Statistics.}}
\newblock Springer-Verlag New York, 2002.

\bibitem{Pennylane}
V.~Bergholm~\textit{et al}.
\newblock {PennyLane: Automatic Differentiation of Hybrid Quantum Classical
  Computations.}
\newblock {\em arXiv:1811.04968}, 2018.

\end{thebibliography}
\bibliographystyle{unsrt}

\end{document}